\documentclass[aps,prd,twocolumn,nofootinbib,superscriptaddress,showpacs]{revtex4}
\usepackage{amssymb,graphics,graphicx, epsfig}
\usepackage{graphicx}% Include figure files
\usepackage{amsfonts}
\usepackage{amssymb}
\usepackage{bm}% bold math
\usepackage{color}
\newcommand{\beqn}{\begin{eqnarray}}
\newcommand{\eeqn}{\end{eqnarray}}
\newcommand{\be}{\begin{equation}}
\newcommand{\ee}{\end{equation}}

\def\s1{$s_{\alpha}$}
\def\s2{$s_{\gamma}$}
\def\s3{$s_{\delta}$}
\def\c1{$c_{\alpha}$}
\def\c2{$c_{\gamma}$}
\def\c3{$c_{\delta}$}
\def \cha{\widetilde{\chi}^{\pm}_1}
\def \chb{\widetilde{\chi}^{\pm}_2}
\def \na{\widetilde{\chi}^{0}}
\def \nb{\widetilde{\chi}^{0}_2}
\def \nc{\widetilde{\chi}^{0}_3}
\def \nd{\widetilde{\chi}^{0}_4}

\newcommand{\gappeq}{\mathrel{\rlap {\raise.5ex\hbox{$>$}}
{\lower.5ex\hbox{$\sim$}}}}
\newcommand{\lappeq}{\mathrel{\rlap{\raise.5ex\hbox{$<$}}
{\lower.5ex\hbox{$\sim$}}}}

\begin{document}
\noindent
\title{
\begin{flushright}
\scriptsize{
\hspace{5cm}
MCTP-09-48\\
NUB-3264\\
YITP-SB-09-23}
\end{flushright}
Explaining PAMELA  and WMAP  data  through Coannihilations   \\
in Extended SUGRA with Collider Implications  }
%%%%%%%%%%%%%%%%%%%%%%%%%%%%%%%%%%%%%%%%%%%%%%%%%%%%%
\author{Daniel Feldman}
\affiliation{Department of Physics,
Northeastern University,  Boston, MA 02115, USA }
%%%%%%%%%%%%%%%%%%%%%%%%%%%%%%%%%%%%%%%%%%%%%%%%%%%%%
\affiliation{ Address after August 1st:  Michigan Center for Theoretical Physics,
 University of Michigan, Ann Arbor, MI 48109, USA}
%%%%%%%%%%%%%%%%%%%%%%%%%%%%%%%%%%%%%%%%%%%%%%%%%%%%%
\author{Zuowei Liu }\affiliation{C. N. Yang Institute for Theoretical Physics,
\\ Stony Brook University, Stony Brook, NY 11794, USA}
%%%%%%%%%%%%%%%%%%%%%%%%%%%%%%%%%%%%%%%%%%%%%%%%%%%%%
\author{Pran Nath$^{1}$}
\author{Brent D. Nelson$^{1}$}
%%%%%%%%%%%%%%%%%%%%%%%%%%%%%%%%%%%%%%%%%%%%%%%%%%%%%
%%%%%%%%%%%%%%%%%%%%%%%%%%%%%%%%%%%%%%%%%%%%%%%%%%%%%
\begin{abstract}
The PAMELA positron excess is analyzed within the framework of
nonuniversal SUGRA models with an extended $U(1)^n$ gauge symmetry
in the hidden sector leading to neutralino dark matter with either a
mixed Higgsino-wino LSP or an essentially pure wino dominated LSP.
The Higgsino-wino LSP can produce the observed PAMELA positron excess
and satisfy relic density constraints in the extended class of
models due to a near degeneracy of the mass spectrum of the extended
neutralino sector with the LSP mass. The simultaneous satisfaction
of the WMAP relic density  data and the PAMELA data is accomplished
through a co-annihilation mechanism ($B_{\rm Co}-mechanism$), and
leads to predictions of a neutralino and a chargino in the mass 
range (180-200)~GeV 
as well as low lying sparticles accessible at colliders.
We show that the models are consistent with the anti-proton
constraints from PAMELA as well as the photon flux data from EGRET 
and FERMI-LAT. Predictions for the scalar neutralino proton cross
section relevant for the direct detection of dark matter are also
discussed and signatures at the LHC for these PAMELA inspired models
are analyzed. It is shown that the mixed Higgsino-wino LSP model
will be discoverable with as little as 1~fb$^{-1}$ of data 
and is thus a prime candidate for discovery in the low
luminosity runs at the LHC.

\end{abstract}
\maketitle
 \preprint{YITP-SB-09-23}
\section{Introduction}
The recent results from the PAMELA experiment~\cite{pamela}
which indicate a large excess in positron production in the
galactic halo, along with a lack of any significant excess
in the  anti-proton flux in the same experiment, have resulted
in a large effort to understand the data. Further, the results
from the FERMI-LAT experiment~\cite{Abdo:2009zk} relating to
the electron and photon fluxes emanating from the galaxy have
also recently been reported and several interpretations have
been put forth. These include particle physics
models~\cite{Cirelli:2008id,kaneetal,Hisano08,Feldman:2008xs,barger,other,decay,susyattempts,photon,neutrinos}
as well as models where the source of the PAMELA excess is of
astrophysical origin~\cite{Yuksel}. Here, we give an analysis of the
positron excess in a supergravity (SUGRA) framework with the neutralino
as the lightest R-parity odd supersymmetric particle (LSP),
where we go beyond the parameter space of the minimal supergravity
grand unification model~\cite{msugra}. As pointed out
in Ref.~\cite{Turner:1989kg}, annihilations of neutralinos into
$W^{+}W^{-}$ and their subsequent decay can provide a positron
excess. One outstanding issue in supersymmetric interpretations
of the positron excess has been that models with large neutralino
annihilation rates into $W^{+}W^{-}$  states tend to imply
a thermally-produced cold dark matter relic density which is as
much as 100~times smaller than the amount indicated by the recent 
WMAP data~\cite{WMAP}. This problem has been discussed in the context
of a non-thermal mechanism which is capable of generating the right
amount of the relic density~\cite{kaneetal,Hisano08}. Such non-thermal
mechanisms could take place via moduli decay
which can arise in various softly broken supersymmetric
theories~\cite{Moroi:1999zb}.

In this paper we consider an alternative mechanism that can generate a
relic density for the LSP neutralino which utilizes the idea of
coannihilation of the LSP with a cluster of states in the hidden
sector which are degenerate in mass with the LSP.
Phenomena of this type can arise in a broad class of models
including extended SUGRA models with a Fayet-Iliopoulos term, models
with kinetic mixings, models with  the Stueckelberg mechanism, and
in string and D-brane models. In many such models connector fields
exist which have non-trivial gauge transformations under the hidden
as well as under the visible sector gauge groups and allow for 
mixings of the LSP with the hidden sector gauginos and chiral
fermions. The mixings between the visible and the hidden sector are
typically constrained by the precision electroweak data.
There are many examples of models of this type that can be constructed,
but here we will consider one specific concrete manifestation as a  
representative of this  class of models. More specifically, we
will assume that there is a set of connector fields (axions)  that transform
non-trivially under the hidden sector  $U(1)_X^n$ gauge group
({\em i.e.}, the product $U(1)_X\times U(1)_{X'}\cdot\cdot\cdot$)  as well
as under the hypercharge gauge group $U(1)_Y$. In addition to the
above, in the visible sector we consider a class of extended SUGRA models
with nonuniversalities in the gaugino masses (see, {\em
e.g.},~\cite{NU,NUSGURA,WinoSGURA,NUSGURAColliders} and references quoted therein)
such that the gaugino masses at the scale of grand unification are of the form 
${\tilde m}_i=(1+\delta_i)m_{1/2}$, $i=1,2,3$ for the gauge groups $U(1)_Y,
SU(2)_L, SU(3)_C$. 
Within the above framework we discuss
illustrative model points: one of these leads to an LSP which has a
mixed Higgsino-wino content while the other is essentially purely
wino. It is found that the model with the mixed Higgsino-wino
content can produce the observed positron excess and at the same time
can coannihilate with the hidden sector gauge and chiral fermions to
produce a relic density consistent with the WMAP~\cite{WMAP} constraints.
 In contrast, for
the model where the LSP is essentially pure wino, the above phemomenon fails
to bring the relic density within the reach of the WMAP data and thus a  nonthermal
mechanism is needed to generate the right relic density. We give now
the details of the analysis.

\noindent
\section{ The General $B_{\rm Co}$  Mechanism}
As already noted in the preceding section, a fit to the PAMELA data
with annihilating dark matter requires a relatively large
annihilation cross section in the halo which is as much as two
orders of magnitude larger (for di-boson production) than the
thermal cross section needed to fit the data on the relic density
of cold dark matter (CDM) consistent with WMAP. The main mechanisms
discussed in the literature to reconcile the two include the
enhancement of the velocity averaged annihilation cross section
$\langle \sigma v \rangle$ in the halo either by annihilation near a
Breit-Wigner pole~\cite{Feldman:2008xs} or by non-perturbative
enhancements (see, e.g., \cite{Hisano:2003ec,Cassel:2009pu} and
references therein).  

Here we will consider an alternative mechanism consisting of a set of particles nearly degenerate in mass with the LSP but which have suppressed interactions 
%with the visible sector states of the MSSM.
with the visible sector states whose particle content
is that of the minimal supersymmetric extension of the
standard model (MSSM).
 When such a hidden sector is present it can act as a reservoir, significantly boosting 
 the resulting thermal relic density of the LSP. 
Here we have in mind a set of hidden sector states acting as a reservoir, similar in spirit to the works of Ref.~\cite{Feldman:2006wd} (see also \cite{Profumo:2006bx}).
Additionally some states in the MSSM sector may also be degenerate in mass with the LSP.
In fact for our specific models we will find that the light chargino of the MSSM will also be degenerate with the LSP.
Thus,  consider
$n_v$ number of sparticles in the visible sector which are essentially degenerate
and $n_h$ number of states in the hidden sector that are essentially  degenerate with the
neutralino LSP  but have
suppressed interactions relative to the MSSM interactions.
 In general
the relic density is governed by  $\sigma_{\rm eff}=\sum_{A,B}
\sigma_{AB}\gamma_A\gamma_B\ $,  where the $\gamma_A$ are the
Boltzmann suppression factors~\cite{Griest:1990kh} \beqn \gamma_A=
\frac{g_A(1+\Delta_A)^{3/2} e^{-\Delta_A x}} {\sum_B g_B
(1+\Delta_B)^{3/2}e^{-\Delta_B x}}\, . \label{gamma}
 \eeqn
Here $g_A$ are the degrees of freedom of $\chi_A$,
$x={m_{\tilde\chi^0}}/{T_x}$ with ${T_x}$ the temperature and $\Delta_A
=(m_{\chi_A}-m_{\tilde\chi^{0}})/m_{\tilde\chi^0}$, and $m_{\tilde\chi^0}$ is
the mass of the LSP ($\tilde \chi^0$). The relic abundance of dark matter at current
temperatures obeys the well known proportionality $ \Omega_{\tilde\chi^0} h^2  \propto
\left[\int_{x_f}^{\infty} \langle \sigma_{\rm eff} v\rangle
\frac{dx}{x^2}\right]^{-1}\ $ where $x_f$ is the value of $x$ at
freeze-out. Because of the small couplings of the hidden sector to
the visible sector, one has $\sigma_{a\alpha},
\sigma_{\alpha\beta} ~{\ll}~ \sigma_{ab}$, where $a,b=1,\ldots,n_v$ label
the  MSSM sector particles that are essentially degenerate with the LSP and
$\alpha, \beta=1,\ldots,n_h$ label the hidden sector particles that
are also essentially degenerate with themselves and the LSP. In this approximation 
one has
 \beqn
 \Omega_{\rm \tilde\chi^0} h^2 \simeq B_{\rm Co} (\Omega_{\tilde\chi^0} h^2)_{\rm MSSM},
 \label{3.5}
 \eeqn
where $(\Omega_{\tilde\chi^0} h^2)_{\rm MSSM}$ is the relic density
as canonically calculated using, for example, the standard tools~\cite{Gondolo:2004sc} and
$\Omega_{\tilde\chi^0} h^2$ is the true relic density when
coannihilation
effects from the hidden sector are taken
into account. Further,  $B_{\rm Co}$ is the enhancement or the boost to
the relic density that comes from effects of coannihilation in the
hidden sector.  With $\sigma_{a \alpha}, \sigma_{\alpha \beta} \ll \sigma_{ab}$ , $B_{\rm Co}$ is then given by
\beqn
 B_{\rm Co} \simeq  \frac{\sum_{a,b}\int_{x_f}^{\infty} \langle \sigma_{ab}
 v \rangle \gamma_a \gamma_b \frac{dx}{x^2}}{\sum_{a,b}\int_{x_f}^{\infty} \langle
 \sigma_{ab}  v \rangle \tilde \gamma_a \tilde\gamma_b \frac{dx}{x^2}},\nonumber \\
 \gamma_a=\frac{g_a(1+\Delta_a)^{3/2} e^{-\Delta_a x}}
 {\sum_{b} g_b(1+\Delta_b)^{3/2} e^{-\Delta_b x}}, \nonumber \\
 \tilde\gamma_a=\frac{g_a(1+\Delta_a)^{3/2} e^{-\Delta_a x}}
 {\sum_{A} g_A(1+\Delta_A)^{3/2} e^{-\Delta_A x}}. 
 \eeqn
Here $A$ runs over channels which coannihilate both in the MSSM sector
and in the hidden sector ({\em i.e.}, $A$ =1,..,$n_v+n_h$) and $g_A$
are the degrees of freedom for particle~$A$; for example, $g=2$ for a neutralino
and $g=4$ for a chargino. 
In the limit
 that $(\sigma v)_{ab}$  are
independent of $a,b$ and all
 $\Delta_A$ nearly vanish, we find the simple relation
%%%%
\beqn
 B_{\rm Co}\simeq (1+ \frac{d_h}{d_v})^2\, .
 \label{3.6}
\eeqn
Here $d_h=\sum_{\alpha} g_{\alpha}$ is the number of degrees of
freedom for the coannihilating channels in the hidden sector (with
suppressed cross sections in the coannihilation process) and $d_v$
is the number of degrees of freedom in the MSSM sector for the coannihilating
channels which contribute to the coannihilations with the LSP and have interactions
of normal strength. For the  $U(1)^n$ hidden sector model 
with each $U(1)$ providing two Majorana states 
with the chargino
coannihilating with the LSP, 
under conditions of essentially complete degeneracy of the chargino and the LSP\footnote{This value
of $B_{\rm Co}$  is the asymptotic limit when there is a complete
degeneracy of the matter in the hidden sector and of the chargino
with the LSP. There is also an upper asymptotic limit which
corresponds to a complete split of the chargino from the LSP which
gives $B_{\rm Co}=(1+2n)^2$, so as the chargino moves from a
complete degeneracy with the LSP to a complete split, one goes from
$(1+\frac{2}{3}n)^2\to (1+2n)^2$. For $n=3$ the above corresponds to
the transition: $B_{\rm Co}=9\to B_{\rm Co}=49$.}, $B_{\rm Co}=(1+\frac{2}{3}n)^2$, while as this degenercy becomes lifted, 
the maximal  value is
 $B_{\rm Co} =(1+2n)^2$.
\noindent
\section{SUSY models with Enhancement of the thermal relic density via  $\bf B_{\rm Co}$}
\label{sec:models}
As noted above
a~$B_{\rm Co}$ of the type discussed in Eq.~(\ref{3.6}) can be
obtained in a  class of  models  where the hidden sector has
suppressed interactions with the visible  sector. Below we construct
one explicit example where a hidden sector with an extended $U(1)^n$
gauge symmetry couples with the MSSM sector via the hypercharge
gauge field. Specifically, we consider an additional
contribution to the MSSM to be of the form~\cite{Kors:2004dx},
\cite{Feldman:2006wd}
\beqn \Delta\mathcal {L} = \int d^2\theta d^2\bar \theta
\sum^{N_S}_{m=1} \left[\sum^{N_V}_{l=1} M_{l,m}V_l +  (\Phi_{m}+\bar
\Phi_{m}) \right]^2, \label{2.9} \eeqn
where $V=\{B,X,X',X'' \ldots \}$ are vector
supermultiplets which include the hypercharge gauge multiplet $B$, and
$\Phi_m$ are a collection of chiral supermultiplets and
$(N_S,N_V>N_S)$ are the number of (axions,vectors). 

 Thus for the $U(1)^n$ extended
models we consider $N_S=n$ and $N_V=n+1$
such that $N_S$
number of  vector bosons absorb $N_S$ number of axions leading to
$N_S$ number of massive $Z'$
bosons~\cite{Kors:2004dx,FLNSt}. A full analysis would include also the electroweak 
symmetry breaking from the MSSM sector generating a mixing between the hidden 
and the visible sectors.
 Models of this type can
arise in a very broad class of theories including extensions of the
SM, extended supergravity models, and in strings and in D-brane
models.
  Specifically we will be interested in an extension of supergravity models  with the hidden sector gauge
group $U(1)_X^n$ with mixings between the visible and the hidden
sector given by Eq.(\ref{2.9}). This extension then leads to a $(4+2n)\times
(4+2n)$ dimensional neutralino mass matrix of the form
%
%%%%%%%%%%%%%
\beqn {\mathcal M}^{[1/2]} =
\left( \begin{array}{c|c}
[{\bf M_1}]_{2n \times 2n} & [{\bf M_2}]_{2n \times 4}  \\ \hline
[{\bf M_2}]^T_{4 \times 2n}  & [{\bf S}]_{4\times 4}  \\
\end{array} \right), 
\label{2.8a}
\eeqn
where 
\beqn
{\bf S}_{4\times 4} =\left( \begin{array}{cccc}
\tilde m_1 & 0 & \alpha & \beta\\
0 & \tilde m_2&  \gamma & \delta \\
 \alpha &  \gamma & 0 & -\mu \\
 \beta  & \delta &  -\mu & 0
\end{array} \right),\label{2.8}\eeqn
%%%%%%%%%%%%%
and where $\alpha= -c_{\beta}s_{W}M_Z$, $\beta= 
s_{\beta}s_{W}M_Z$, $\gamma =c_{\beta} c_{W} M_Z$, $\delta= -s_{\beta}c_{W}M_Z$. Here
($s_{\beta}, c_{\beta}) =(\sin\beta,\cos\beta)$, where
$\tan\beta= \langle H_2 \rangle / \langle H_1 \rangle $ and $\langle H_2\rangle$ gives mass to the up quarks and
$\langle H_1 \rangle$ gives mass to the down quarks and the leptons, and $(s_W,
c_W)=(\sin \theta_W, \cos\theta_W)$, where $\theta_W$ is the weak
angle. The remaining quantities in Eq.(\ref{2.8a}) are
\begin{eqnarray} \left[\mathbf{M}_1\right]_{2n \times 2n} &=& \left( \begin{array}{cccc}
 M^{(n)}_1 \hat U & \hat O & \ldots & \hat O  \\
\hat O    &  M^{(n-1)}_1 \hat U   & & \hat O \\
\vdots    &     & \ddots & \vdots \\
\hat O  & \hat O &\ldots &  M^{(1)}_1 \hat U
\end{array} \right),
\label{2.8b}
\end{eqnarray}
and 

\begin{eqnarray}
\left[\mathbf{M}_2\right]_{2n \times 4} &=& \left(
\begin{array}{ccc}
 M^{(n)}_2 \hat P & \hat O  \\
 M^{(n-1)}_2 \hat P & \hat O  \\
\vdots    & \vdots     \\
 M^{(1)}_2 \hat P & \hat O 
\end{array} \right),
\label{2.8c}
\end{eqnarray}

%\begin{eqnarray}
%\left[\mathbf{M}_2\right]_{2n \times 4} &=& \left(
%\begin{array}{ccccc}
% M^{(n)}_2 \hat P & \hat O & \ldots & \hat O \\
% M^{(n-1)}_2 \hat P & \hat O & \ldots &   \hat O \\
%\vdots    & \vdots    & \ddots & \vdots \\
% M^{(1)}_2 \hat P & \hat O & \ldots &   \hat O
%\end{array} \right) 
%\label{2.8c}
%\end{eqnarray}

where the ellipses indicate additional $U(1)$s and where
%%%%%%%%%%%%%%%%
\beqn \hat U =
\left( \begin{array}{cc}
0 & 1\\
1 & 0
\end{array} \right),~~
 \hat P =
\left( \begin{array}{cc}
1 &0\\
0 & 0
\end{array} \right),~~
 \hat O =
\left( \begin{array}{cc}
0 & 0\\
0 & 0
\end{array} \right).
\eeqn

The $4\times 4$ dimensional matrix of Eq.(\ref{2.8})
  is the neutralino mass  matrix in the MSSM sector, the $(2n)\times (2n)$ dimensional
 mass matrix of Eq.(\ref{2.8b})
 is the neutralino mass matrix in the
hidden sector, and the $2n\times 4$ dimensional matrix of  Eq.(\ref{2.8c})
is the matrix of 
 off-diagonal terms which produce the mixings
between the hidden sector and the MSSM sector.
 These mixings are
controlled by the ratios $\epsilon\equiv M_2/M_1$, etc. all of which we assume to be much smaller than unity~\cite{FLNSt}. We also assume that
the hidden sector neutralinos are all essentially degenerate with the LSP
neutralino. Thus in the diagonal basis we have the set of neutralino
states $\tilde\chi^0_i$ ($i=1$,$\ldots$,$4+2n$) where the last $n_h\equiv
2n$ are the set of neutralinos from the hidden sector. The above
serves as an illustrative example, but our analysis would apply to
any class of models which satisfy the conditions discussed in the
beginning of this section.

We discuss now two specific model points which fit the PAMELA data
but lead to significantly different sparticle spectra and signatures
in the direct detection of dark matter and at colliders. \\

\noindent
 {\bf  Higgsino-wino model (HWM)}\\
Here the soft  nonuniversal SUGRA  parameters are \beqn
(m_0,m_{1/2},A_0)=(800,558,0)~{\rm GeV}, \tan\beta=5,  \nonumber \\
{\rm sign}(\mu)=+, ~\delta_{1,2,3}=(-.09,-.50,-.51)\, , \,\,
\label{5.4} \eeqn and   the gaugino-Higgsino content of the LSP is
$\tilde\chi^0 = 0.726 \lambda_B -0.616\lambda_W +0.260 \tilde h_1
-0.160\tilde h_2+  C_h\Delta\tilde\chi^0_h$. Thus the LSP has a strong
Higgsino component in addition to strong wino and bino components. Nevertheless
we will refer to this model as the Higgsino-wino model (HWM) because these components 
play a major role in the analysis to follow.
The quantity $C_h \Delta \tilde\chi^0_h$ is the component in the hidden sector
and is found by excplicit calculation to be rather small ($|C_h|<1\%$ for $n=3$).\\

\noindent
 {\bf Pure wino model (PWM)}\\
  Here the soft parameters  are
\beqn (m_0,m_{1/2},A_0)=(1000,850,0)~{\rm GeV}, \tan\beta=10,
\nonumber \\ {\rm sign}(\mu)=+, ~\delta_{1,2,3}=(0,-.7,0)\, ,\qquad
\qquad \label{5.1} \eeqn and the gaugino-Higgsino content of the LSP
is $\tilde\chi^0 = 0.009 \lambda_B -0.996\lambda_W +0.081 \tilde h_1
-0.023\tilde h_2 + C_h \Delta\tilde  \chi^0_h$. 
In this model
the LSP is dominated by
the wino component and thus this model will be referred to as
the (essentially) pure wino model (PWM), while the component in the hidden sector $C_h \Delta
\tilde\chi^0_h$ is still small (i.e., also $|C_h|<1\%$, for $n=3$).\\

For both model points given above one finds the following mass
relations \beqn m_{\tilde\chi^{\pm}}\simeq m_{\tilde\chi^0},
~m_{\tilde\chi_2^{\pm}}\simeq m_{\tilde\chi_3^0}\simeq m_{\tilde\chi_4^0}\, .
\label{5.6} \eeqn These mass relations follow simply from the
condition that $\tilde m_2 < \tilde m_1$, 
$|\mu| >> M_Z$.
 However, a distinguishing feature of the HWM is that the mass gap between
the LSP and the chargino is order $\sim 10$ GeV while for the case of the PWM
the mass gap is order the pion mass. 
The neutralino annihilation in
  both  models
is dominated by $W^{+}W^{-}$ production in the halo. 
We will discuss this in much more
detail in what follows.
\noindent
\section{The PAMELA Positron Data and Thermal Neutralino Dark Matter }
\label{sec:positron}
One issue of
 interest in this work is to address the compatibility
of the recent PAMELA positron data and the WMAP data. Thus, we begin with a
brief review of the calculation of the primary positron flux relevant for
the models considered here. 

The positron flux which enters as a solution to the diffusion loss
equation under steady state conditions is given
by~\cite{Hisano:2005ec},\cite{Delahaye:2007fr},\cite{Cirelli:2008id}
\beqn \Phi_{\bar e}(E) =   \frac{ B _{\bar e} v_{\bar e}}{8\pi
b(E)} \, \frac{\rho^2_\odot}{m^2_{\tilde\chi^0}} F(E)~, 
\label{4.1}
\eeqn 
\beqn
 F(E) = \int_{E}^{M_{\tilde\chi^0}} dE' \sum_{k} \langle \sigma v\rangle^k_{\rm halo}
 \frac{dN^{k}_{\bar e}}{dE'} \cdot \mathcal{I} (E,E').
\label{halo_posit_flux} \eeqn
In the above, $B_{\bar e}$ is a so-called boost
factor which parameterizes the possible local inhomogeneities of the
dark matter distribution.
Large boost factors have been used in the literature, even as large
as 10,000, to explain the PAMELA data. However, we will show that
there are various cases where boost factors in the range $\sim
(1-5)$ fit the PAMELA data in  models of supersymmetry (SUSY). In
Eq.~(\ref{4.1}) $v_{\bar e}$ is the positron velocity,
$v_{\bar e} \sim c$, ${\rho}_{\odot} =\rho(r_{\odot})$ is the local
dark matter density in the halo (with $r_{\odot} \sim 8.5~\rm kpc$)
and ${\rho}_{\odot}$ lies in the range $(0.2-0.7) ~\rm
GeV/cm^3 $~\cite{Kamionkowski:2008vw}. Further, $b(E)$ in
Eq.(\ref{4.1}) is given
by $b(E) = E_0 (E/E_0)^2
/\tau_E$, with $E$ in  GeV and $E_0 \equiv 1 ~\rm GeV$, and where
$\tau_E = \tau ~10^{16} ~{\rm s}$, where $\tau$ values as large (1-5)
have been considered in the
literature and we adopt $\tau =3$ \cite{Longair:1994wu} . Here $\langle \sigma v \rangle_{
\rm halo} $ is the velocity averaged cross section in the halo
of the galaxy and  
${\cal I }(E,E')$ is the halo function. 
We have considered both the Navarro,
Frenk and White~(NFW) and Moore
et.~al~\cite{Navarro:1996gjMoore:1999gc} profiles coupled with
various diffusion models. The diffusion loss equation depends on the
propagation parameters $\delta$, $K_0$ solved in the region modeled
with cylindrical symmetry bounding the galactic plane with height
$2L$. The dimensionless halo function has been parameterized to
satisfy constraints on the boron to carbon
ratio~\cite{Delahaye:2007fr,Cirelli:2008id} such that
\beqn
\mathcal{I}(E,E')  = a_0 + a_1 \tanh\left(\frac{b_1-\ell}{c_1}\right) \times \\
\nonumber \left[ a_2 \exp \left( -\frac{(\ell - b_2)^2}{c_2}\right)
+ a_3 \right] \eeqn
and where $\ell = \log_{10}\lambda_D $ with $\lambda_D$ in units of
kpc \beqn \lambda_D (E,E') =2\sqrt{  K_0\, \tau_E\, D^{-1} \,
(E^{-D}-E'^{-D})} \\ D= 1-\delta. \label{halo_fit_positrons} \eeqn
%
%%%%%%%%%%%%%%%%%%%%%%%%%%%%%%%%%%%%%%%%%%%%%%%%%%%%%%%
% ------------------ TABLE I --------------------------
\begin{table}[h]
\begin{center}
\begin{tabular}{|lccc|}
\hline
Model  & $\delta$ & $K_0$ (${\rm kpc}^2/{\rm Myr}$) & $L$ ( kpc)  \\
\hline
MIN(M2)  & 0.55 &  0.00595 & 1 \\\hline
\end{tabular}
\begin{tabular}{|c | c |c|c|c|c|c|c|c|c|}\hline \nonumber
\rm{Halo model}  \rm{propagation} & $a_0$ & $a_1$ & $a_2$ & $a_3$\\
\rm{NFW} \rm{MIN(M2)} & 0.500 & 0.774 & -0.448 & 0.649\\
{\rm Halo function coeff.} & $b_1$ & $b_2$ & $c_1$ & $c_2$\\
& 0.096 & 192.8 & 0.211 & 33.88 \\
\hline
\end{tabular}
\begin{tabular}{|cccc|}
\hline \nonumber
 $w^{\bar e}_0$ & $w^{\bar e}_1$ &$ w^{\bar e}_2 $& $w^{\bar e}_3$  \\  \hline
   - 2.28838& -0.605364 &- 0.287614 &-0.762714  \\
   \hline
$w^{\bar e}_4$& $w^{\bar e}_5$ & $w^{\bar e}_6$ & $w^{\bar e}_7$
\\  \hline
- 0.319561 & -0.0583274 & - 0.00503555 & -0.00016691\\
\hline
\end{tabular}
\caption{ The class of models consistent with the
boron/carbon ratio~\cite{Delahaye:2007fr} considered here. The MED and MAX halo
models are not shown since they tend to overproduce the positron
flux at low energies. Also shown are the coefficients $(a_i,b_i)$
entering in the dimensionless halo function for positron
propagation~\cite{Cirelli:2008id} where the NFW profile is used and
the MIN(M2) diffusion parameters enter. The last set of entries are
the fragmentation coefficients $w^{\bar e}_i$ for the $W^+W^-$
mode~\cite{Hisano:2005ec}. } \label{postab}
\end{center}\end{table}
% ------------------ End table --------------------------

For the case of dark matter in the halo which primarily annihilates
into $W$~boson pairs, the positron fragmentation function can be
expressed as~\cite{Hisano:2005ec}
\beqn
\frac{d N^{WW}_{\bar e}}{dx}
 =\exp\left[\sum_{n} w^{\bar e}_n (\ln(x))^n\right],
\eeqn where $x = E/m$ and the fragmentation function was fit in the
analysis of~\cite{Hisano:2005ec} with HERWIG~\cite{Corcella:2000bw}.

The positron fluxes in the absence of dark matter annihilations have
been parameterized in~\cite{Moskalenko:1997gh} using fits to the analysis
which we adopt here.  
For the case of positrons annihilating
strongly into $W$~bosons, we find that the MED and MAX models with NFW
or Moore profile tend to overproduce the positron flux data at
energies well above the region of solar modulation. Thus the class of
models we consider restricts the profile/diffusion model to a MIN(M2) scenario, at least for the case of positrons. We thus display
only the fit parameters for this scenario in
Table (\ref{postab}). The fit parameters for the halo
function are also shown in Table (\ref{postab}) along with
the fragmentation functions for dark matter annihilations into $W^+W^-$
states.
%%%%%%%%%%%%%%%%%%%%%%%%%%%%%%%%%%%%%
%%  FIGURE 1
%%%%%%%%%%%%%%%%%%%%%%%%%%%%%%%%%%%%%
\begin{figure}[t]
  \begin{center}
             \includegraphics[width=8.1cm,height=7cm]{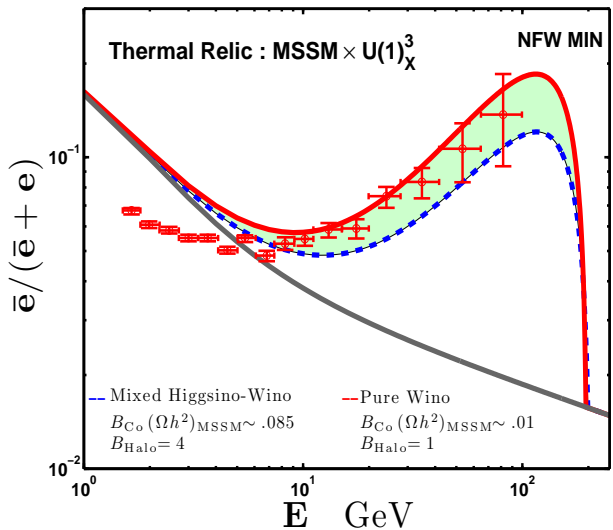}
    \caption{ SUSY predictions for the positron flux ratio:
The top curve is for the PWM (see Eq.~(\ref{5.1})) where the
neutralino is wino dominated, with $m_{\na} =199$~GeV, and $\langle
\sigma v \rangle_{WW} = 1.96 \times 10^{-24} \rm cm^3/s$ with a
$B_{\rm clump}$ of $1.0$ and $\Omega h^2 \sim .01$ with a $B_{\rm
Co}$ $\sim 9$ (the asymptotic value for the wino case). The middle
curve is for the HWM (see Eq.~(\ref{5.4})) where the neutralino has
a mixed wino, bino and Higgsino content, with $m_{\na} =195$~GeV,
and $\langle \sigma v \rangle_{WW} = 0.28 \times 10^{-24} \rm
cm^3/s$ with $B_{\rm clump}$ of $4$ and gives rise to an $\Omega h^2
= .085$, with a $B_{\rm Co}$ of 16. We have taken $\rho_{\odot}~ =
0.6 ~ {\rm GeV/cm^3}$ and $\tau =
3$ for both curves. Slightly larger clump
factors can easily accommodate a downward shift in the product
$\rho^2_{\odot}\tau_{E}$. The bottom curve is the background. The
experimental data is from~\cite{pamela}.
}
\label{pamfig}
  \end{center}
\end{figure}
%
%%%%%%%%%%%%%%%%%%%%%%%%%%%%%%%%%%%%%
%%  FIGURE 2
%%%%%%%%%%%%%%%%%%%%%%%%%%%%%%%%%%%%%
\begin{figure}[t]
  \begin{center}
     \includegraphics[width=8.1cm,height=7cm]{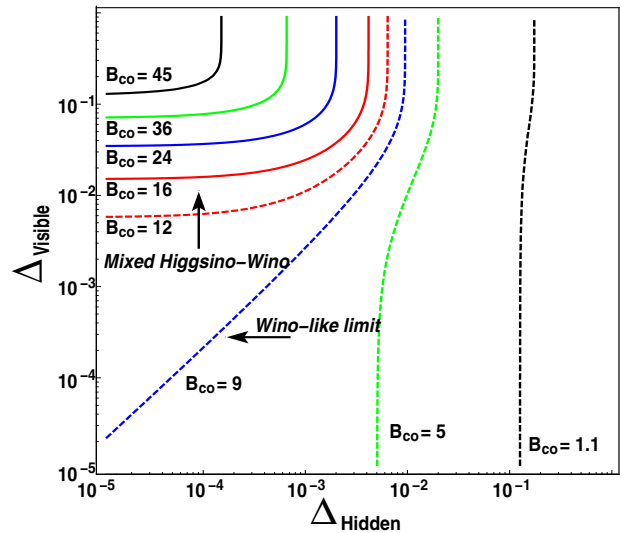}
    \caption{
An illustration of the $B_{\rm Co}$-mechanism with a contour plot of
constant $B_{\rm Co}$ in the $\Delta_{\rm Visible}$ vs $\Delta_{\rm
Hidden}$ plane for the $U(1)^3$ extended SUGRA model. 
Large values of $B_{\rm Co}$ are admissible for the
mixed Higgsino-wino case (see the text).  The region to the right of the
contour $B_{\rm Co}=9$ is the domain of the PWM  while
 the  region to the left of it is the domain of the HWM. Thus the
 largest value of $B_{\rm Co}$  in the PWM is 9 while for the HWM case $B_{\rm Co}$  can be  significantly larger. 
}
\label{DeltaFig}
\end{center}
\end{figure}
As discussed in the previous section, in the analysis here we fix the soft parameters of the SUGRA 
models at the scale of grand unification and examine the implications of the
model at the electroweak scale.  The analysis shows that the mixed
Higgsino-wino model (HWM) can have mass splittings between the LSP and the
chargino of around 10~GeV and still produce relatively large halo
cross sections that do not require large boost factors to fit the
PAMELA data. On the other hand for the pure (or nearly pure) wino
LSP one finds that the mass splitting of the LSP neutralino and the
light chargino is negligibly small and this result is generally a
necessary consequence of the wino dominated class of models. A
numerical analysis of the positron excess for comparison with the
PAMELA experiment is presented in Fig.~(\ref{pamfig}) for the HWM and for the PWM. It is found that both
models are capable of fitting the PAMELA data. A wino LSP of a mass near 200~GeV has been suggested in 
Ref.~\cite{kaneetal} as a good candidate for explaining the positron excess using a methodology and choice of parameters similar to the ones used here, though the model in that study and this study are quite different. 

While both the HWM and PWM are successful at reproducing the PAMELA positron excess, the thermal relic
densities for the two models are very different. For the case of the HWM, the relic density can be consistent with the WMAP data for a thermal LSP.
However, the relic density for the PWM from thermal processes is at best about a factor of~10 too small and here one needs a
non-thermal mechanism to be compatible with WMAP, as was assumed in the work of Ref.~\cite{kaneetal}.  The compatibility
of the relic density with WMAP for the HWM comes about because of
the enhancement from the hidden sector, {\em i.e.}, because of the
$B_{\rm Co}$ factor. As seen in Fig.(\ref{DeltaFig}), $B_{\rm Co}$ can acquire values as large
as~45. This limiting value of $B_{\rm Co}$ is realized only when
there is a complete split between the LSP and the light chargino.
While this value is not achieved in the HWM, the presence of  a
$\sim$ 10~GeV split between the LSP and the chargino does conspire to
produce a $B_{\rm Co}$ in the range $16-20$ while for the PWM the
maximal value one may have is $B_{\rm Co} \sim 9$. While a factor
of~$16-20$ is good enough to boost the relic density for the HWM to
be consistent with WMAP, for the PWM the relic density still falls below the WMAP
corridor and would require a very large collection of degenerate states 
(e.g., $n=11$) to 
bring the relic density within the corridor.

\noindent
\section{Analysis of the $\bar p$ flux}
Quite similar to the analysis
of the positron flux, the anti-proton flux  assumes 
a steady state solution to a diffusion equation, retaining
a factorized form split between the particle physics content and the astrophysical details, the latter
encoding the halo  and the propagation model. 
Thus the  anti-proton flux from annihilating neutralinos in the galaxy is given
by~\cite{Bottino:1994xs}
\beqn \Phi_{\bar p}(T) = B_{\bar p} \,  \frac{v_{\bar p}}{8\pi}
\,  \frac{\rho^2_\odot}{m^2_{\tilde\chi^0}} \, R(T) \, \sum_k \langle
\sigma v\rangle^k_{\rm halo}  \, \frac{dN^k_{\bar p}}{dT} \eeqn
where $T$ is the kinetic energy, and  $R(T)$ has been fit as in Ref.~\cite{Cirelli:2008id} for
various profile/diffusion models. Specifically for the NFW models
one has
\begin{eqnarray}
{\rm MIN}&:& \, {\rm log}_{10}\left[R(T)/{\rm Myr}\right] =
0.913+ 0.601\, r \nonumber \\
 & & -0.309\, r^2 - 0.036\,  r^3+ 0.0122\, r^4  \nonumber \\
{\rm MED}&:& \, {\rm log}_{10}\left[R(T)/{\rm Myr}\right] =
1.860 + 0.517\, r \nonumber \\
 & & -0.293\, r^2 -0.0089\, r^3   + 0.0070\, r^4 \nonumber \\
{\rm MAX}&:& \, {\rm log}_{10}\left[R(T)/{\rm Myr}\right] = 2.740
-0.127\, r \nonumber \\
& & -0.113\, r^2+ 0.0169\,r^3    -0.0009\, r^4 \,
,\label{antiprotonsR}
\end{eqnarray}
where $r= r(T) = \log_{10}(T/{\rm GeV})$, and where the 
$\bar p$ propagation parameters are given in Table~(\ref{diffanti}).
\begin{table}
\begin{tabular}{ccccccc}
Model  & $\delta$ & $K_0$ (kpc$^2$/Myr) & $L$ (kpc) & $V_{\rm conv}$ (km/s) \\
\hline
{\rm MIN}  & 0.85 &  0.0016 & 1  &13.5\\
{\rm MED}  & 0.70 &  0.0112 & 4  & 12 \\
{\rm MAX}  & 0.46 &  0.0765 & 15 &5
\end{tabular}
\caption{Specific models considered here, consistent with the boron/carbon ratio~\cite{Maurin:2001sj}}
\label{diffanti}
\end{table}

As discussed already, for the models of interest we study here,
the contribution to $\langle \sigma v \rangle$ in the halo is dominated by
several orders of magnitude from  the annihilations of neutralino dark matter into $W^+W^-$ bosons.
The $\bar p$ fragmentation functions needed for the analysis of the flux  are given
by~\cite{Bergstrom:1999jc},\cite{Hisano:2005ec} as
\begin{equation}
\frac{dN^{WW}_{\bar p}}{dx}
 =
 \left(p_1 x^{p_3} + p_2|\log_{10}x|^{p_4}\right)^{-1}~,
\end{equation}
where here  $x = T/m_{\tilde\chi^0}$ and $T$ is the kinetic energy. The parameters $p_i$ in the above equation
depend on the neutralino mass and are given by
\begin{equation}
 p_i(m)
 =
 \left(a_{i1}m^{a_{i2}} + a_{i3}m^{a_{i4}}\right)^{-1},~~~~~m=m_{\tilde\chi^0},
\label{pim}
\end{equation}
where the values of $a_{ij}$ are given in Table (III).
\begin{table}[htb]
\begin{center}
  \begin{tabular}{|l|c c c c|}
   \hline
         & $j=1$   & $j=2$   & $j=3$                  & $j=4$\\
   \hline
   $i = 1$ & 306.0 & 0.28  & 7.2$\times 10^{-4}$  & 2.25 \\
   $i = 2$ & 2.32  & 0.05  & 0                    & 0 \\
   $i = 3$ & -8.5  & -0.31 & 0                    & 0 \\
   $i = 4$ & -0.39 & -0.17 & $-2.0\times 10^{-2}$ & 0.23 \\
   \hline
  \end{tabular}
  \caption{\small Coefficients  $a_{ij}$ for
    $W^{+}W^{-}$ process entering into the anti-proton fragmentation functions~\cite{Bergstrom:1999jc}\cite{Hisano:2005ec}.}
\end{center}
 \end{table}
%

%%%%%%%%%%%%%%%%%%%%%%%%%%%%%%%%%%%%%
%%  FIGURE 3
%%%%%%%%%%%%%%%%%%%%%%%%%%%%%%%%%%%%%
\begin{figure}[t] \center
    \includegraphics[width=8.3cm,height=7.4cm]{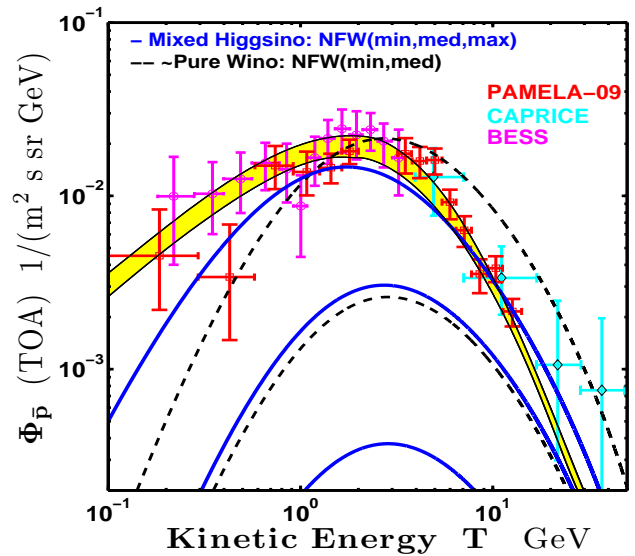}
     \caption{The absolute anti-proton flux for the HWM 
with various halo/diffusion models NFW(MIN,MED,MAX) 
and only (MIN,MED) for the PWM . Here
$\rho_{\odot} = 0.3 ~\rm GeV/cm^3$ (the signal scales as $\rho^2$), $B_{\bar p}=1$. For the
background flux (band) we have adopted the parameterizations of
Ref.~\cite{Bringmann:2006im}. Also shown is the preliminary PAMELA
data~\cite{Bruno,Ricci} along with the earlier data sets from BESS
and CAPRICE~\cite{Orito:1999re}. The analysis here shows that the
models discussed in this work can accommodate the anti-proton flux
constraints.}
  \label{pbarflux}
\end{figure}

The anti-proton flux observed at the top of the atmosphere including
solar modulation can be accounted for by replacing
$\Phi_{\bar{p}}(T)=\Phi_{\bar{p}}(T,\phi_F)
=\Phi_{\bar{p}}\left(T+|Z|\phi_F \right)$ and including a kinetic
energy correction ratio, such that
\beqn 
\Phi^\oplus_{\bar{p}}(T)=\frac{\left((T +m_p)^2 - m_p^2\right)\cdot
\Phi_{\bar{p}}\left(T+|Z|\phi_F \right)}{(T
+|Z|\phi_F + m_p)^2 - m_p^2} 
\eeqn
where $m_p$ is the proton mass, $Z=1$, and the Fisk potential
$\phi_F$ is taken as 500~MV. In Fig.(\ref{pbarflux})
we give an analysis of the anti-proton flux with the parameters
indicated in the caption of the figure 
showing both the SUSY signal and the backgrounds. The analysis
demonstrates that HWM  is only currently constrained
for the MAX diffusion model, while the PWM  is
constrained for a MED diffusion model but is essentially
unconstrained for a MIN diffusion model. 
Similar conclusions regarding the sensitivity of the $\bar p$
flux to the halo/diffusion model have also been made in the first Ref. of  \cite{Feldman:2008xs}.

In summary, the findings here for the PWM are generally in good agreement with 
 the recent results of \cite{kaneetal},
and further we find that  the  HWM
is only very weakly constrained
by the PAMELA anti-proton data as well as by the earlier BESS and CAPRICE data.

\noindent
\section{Photon flux; EGRET AND FERMI-LAT}
The continuum photon source from the annihilating SUSY dark matter
is given by~\cite{Bergstrom:1997fj}
\beqn \frac{d^2\Phi_\gamma}{d\Omega\,dE}=
\frac{1}{2}\,\frac{r_\odot\,\rho^2_{\odot}}{4\,\pi\,m_{\tilde\chi^{0}}^{2}}
\sum_{k} {\langle \sigma v \rangle}^k_{\rm
halo}\frac{dN^k_{\gamma}}{dE}~J(\psi). \eeqn
where $J(\psi) = (r_\odot\rho^2_{\odot})^{-1} \int^{\infty}_0
\rho^2(r) d l(\psi)$ and $r^2=l^2+r^2_\odot-2 l r_\odot \cos{\psi}$
with $\psi$ the angle of integration over the line of sight. After
integration, the astrophysics is encoded in
$\bar{J} \cdot \Delta \Omega$, where $\bar{J} = (1/\Delta \Omega)
\int_{\Delta \Omega} J(\psi)\,d\Omega$ and various values of $ \bar
J$ are given for a collection of angular
maps, in, for example \cite{Cirelli:2009vg}  (see
also~\cite{Berezinsky:1994wva,Fornengo:2004kj},\cite{Strong:2004de}
for early work). As remarked previously, $W$ boson pair production is the dominant
contributor to the photon flux in the models we discuss, and an
upgraded set of fragmentation functions are given
in \cite{Cirelli:2008id} and are shown below
\beqn E\frac{ dN^{\gamma}}{d E} = \exp\left[\sum_n
\frac{w^{\gamma}_n}{n!} \ln^n (E/M)\right]. \eeqn
Separately, effects due to Brehmstrahlung have been studied
in Ref.~\cite{Bergstrom:2005ss} and are easily accounted
for.  Values of $w^{\gamma}_n$ relevant for the
analysis are listed below.
\begin{table}[h]
\hspace{-.2cm}\begin{tabular}{cccccccccc}
 $w^{\gamma}_0$ & $w^{\gamma}_1$ &$ w^{\gamma}_2 $& $w^{\gamma}_3$ & $w^{\gamma}_4$ & $w^{\gamma}_5$ & $w^{\gamma}_6$  \\  \hline
\hline
  -6.751 & -5.741 & -3.514 & -1.964 & -0.8783 & -0.2512 & -0.03369\\
\hline
\end{tabular}
\end{table}
%
%%%%%%%%%%%%%%%%%%%%%%%%%%%%%%%%%%%%%
%%  FIGURE 4
%%%%%%%%%%%%%%%%%%%%%%%%%%%%%%%%%%%%%
\begin{figure}[t]
%\center
% \includegraphics[width=8.2cm,height=7.2cm]{plots/gamma}
 \includegraphics[width=9cm,height=6cm]{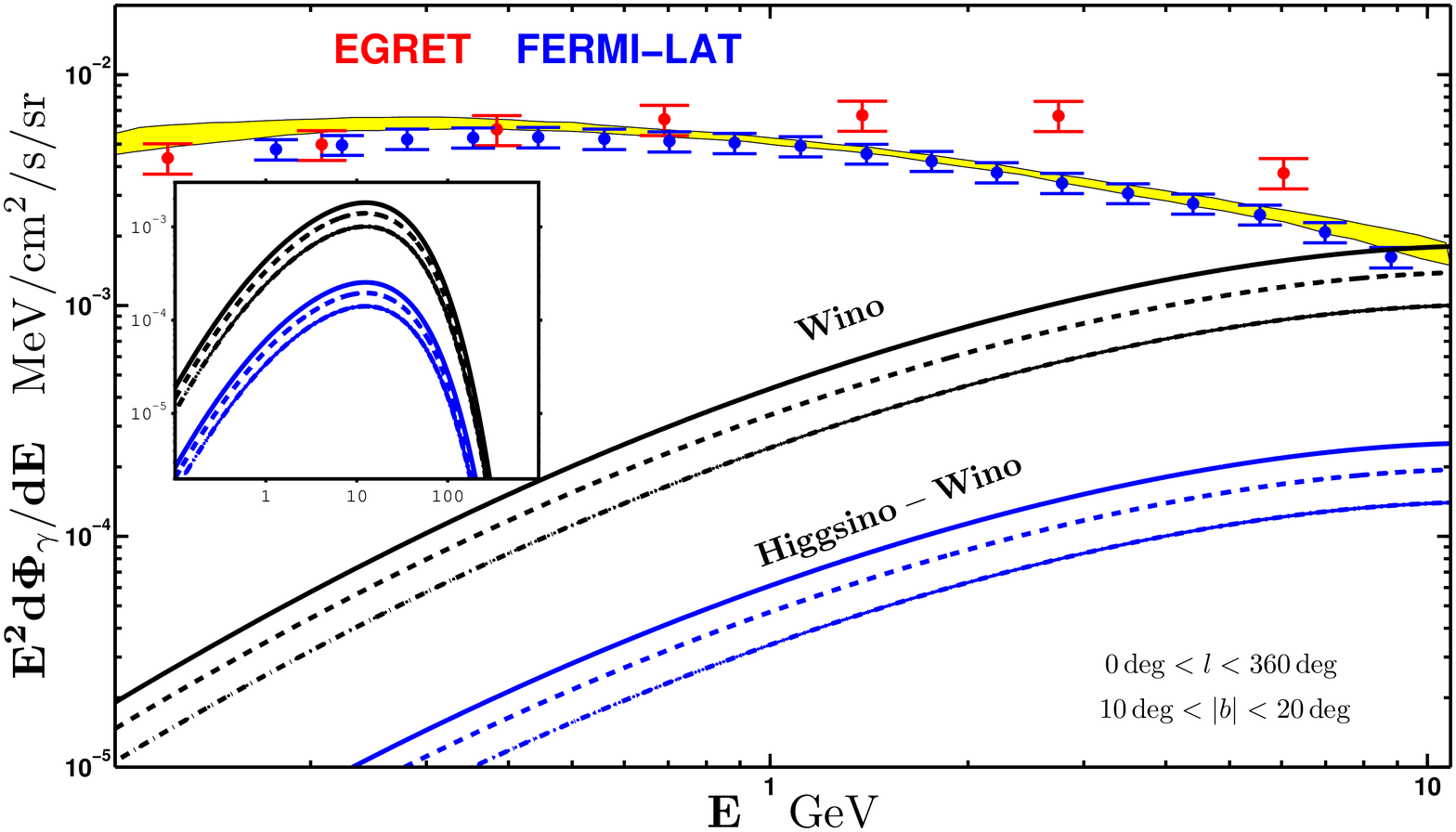}
  \caption{
An exhibition of the gamma ray flux the  HWM and the PWM in the
coordinate range indicated, with the Einasto, NFW, and Isothermal profiles
 with the same halo cross sections as given in 
Figs.(\ref{pamfig},\ref{pbarflux}). Shown are the EGRET~\cite{Hunger:1997we} results
and
FERMI-LAT results as  reported in ~\cite{gig,winer}   along with the background
flux (band). The analysis here shows that both models can
accommodate the photon flux constraint.}
  \label{photonflux}
\end{figure}
\\
Our result for the photon flux is given in Figure~(\ref{photonflux})
where we exhibit the gamma ray flux arising from the annihilation of
neutralino dark matter in the $10-20$~region $(10 \deg < |b| < 20\deg), (0 \deg< l < 360 \deg) $  
[latitude $b$ and longitude $l$]
which is the region relevant for comparison with the preliminary
FERMI-LAT results~\cite{winer}. Figure~(\ref{photonflux}) shows that
the more accurate FERMI-LAT data falls in magnitude below the EGRET
data. Our analysis of the photon flux for the HWM 
is consistent with the FERMI-LAT and EGRET data, while the PWM may show
an excess at extended energy ranges. Note the maximum of the flux for the signal appears at the
last data point of the FERMI data near 10 GeV.

We note that monochromatic sources \cite{Bergstrom:1997fh} (calculated with DarkSusy \cite{Gondolo:2004sc})
 yield a further distinguishing feature between the HWM and the PWM.
Here the PWM will produce a significantly stronger monochromatic source than the HWM which is dictated
by the size of the relative cross sections since $m_{{\tilde \chi}^0}$ are essentially the same for the HWM and PWM:
\begin{eqnarray}
{\rm HWM} &:& \langle \sigma v \rangle^{1-\rm loop}_{\gamma \gamma} = 1.6 \times 10^{-28} ~\rm cm^3/s ,\\\nonumber
{\rm PWM}&:&  \langle \sigma v \rangle^{1-\rm loop}_{\gamma \gamma} = 2.0 \times 10^{-27}~\rm cm^3/s.\\ \nonumber
\\
{\rm HWM}&:&  \langle \sigma v \rangle^{1-\rm loop}_{\gamma Z} = 1.0 \times 10^{-27}~\rm cm^3/s, \\\nonumber
{\rm PWM}&:&  \langle \sigma v \rangle^{1-\rm loop}_{\gamma Z} = 1.3 \times 10^{-26}~\rm cm^3/s.
\end{eqnarray}
Thus while the photon energies will be essentially the same ($m_{{\tilde \chi}^0}$ for the $\gamma \gamma$ channel and
$m_{{\tilde \chi}^0}(1-\delta)$ , $\delta = M^2_Z/(2 m_{{\tilde \chi}^0})^2$ for the $\gamma Z$ channel) in fact, the PWM
would predict a flux an order of magnitude larger than the HWM in both channels. 

\noindent
\section{Effects on Direct Detection}
The direct detection of dark matter is sensitive to the Higgsino content of the LSP.
For the HWM the LSP has a significant Higgsino component and thus
the spin independent cross section for neutralino-proton scattering
in the direct detection dark matter experiments is significant and
may lie within reach of the next generation experiments.
 For the PWM, we remind the reader
 that the LSP is essentially $100\%$ wino, and in this case the spin
independent cross section will be very small, essentially beyond the
limit of sensitivity of near future experiments on the direct
detection of dark matter.  Specifically, a direct calculation using MicrOmegas (see Ref. 2 of \cite{Gondolo:2004sc}) yields a neutralino-proton cross section of
\beqn {\rm HWM}&:& ~\sigma_{\tilde\chi^0 p}^{\rm SI}=6.62\times
10^{-8}\, {\rm pb}, 
\\
{\rm PWM}&:& ~\sigma_{\tilde\chi^0 p}^{\rm SI}=1.15\times 10^{-9}
\,{\rm pb}\, , \eeqn
which translates into a recoil rate on germanium targets of
\beqn {\rm HWM}&:& ~ R= 1.58\times 10^{-2}/{\rm day}/{\rm kg}\, , 
\\
{\rm PWM}&:&~ R= 2.81\times 10^{-4}/{\rm day}/{\rm kg}\, ,
\eeqn
and on liquid xenon targets of
\beqn  {\rm HWM}&:& ~R=2.60 \times 10^{-2}/{\rm day}/{\rm kg}\, ,
\\
{\rm ~PWM}&:&~  R= 4.63 \times 10^{-4}/{\rm day}/{\rm kg}\, ,
\eeqn
when integrated over the entire range of nuclear recoil energies. 
For the pure wino and pure higgsino cases the cross sections can receive
important loop corrections\cite{Hisano:2004pv}. 
From the analysis of \cite{Hisano:2004pv} one can estimate these effects,
and they are found not be significant for the parameter space we investigate.
Clearly, the prospects for the direct detection of the relic LSP  are most promising for the HWM, 
which should be accessible in near future dark matter experiments with an 
improvement in sensitivity by a small factor,
 as is expected to occur.
For the PWM the observation of the spin independent cross section
requires an improvement in sensitivity by a factor of about~100
which is unlikely to happen in the foreseeable future.

%%%%%%%%%%%%%%%%%%%%%%%%%%%%%%%%%%%%%%%%%%%%%%%%%%%%%%%%%%%%%%%%%%%%%%
\noindent
\section{Collider Implications of PAMELA Inspired Models}

%========= table of benchmark phys spectrum ==========
\begin{table}[b]
{\begin{center}
\begin{tabular}{|l|c|c|c|l|c|c|}\cline{1-7}
Mass  & HWM & PWM & & Mass   & HWM & PWM \\
\cline{1-7} $m_{\na}$ & 198.9 & 195.2 &
& $m_{\tilde{t}_{1}}$ & 648.5 & 1516 \\
$m_{\nb}$ & 217.0 & 357.0 &
& $m_{\tilde{t}_{2}}$ & 866.8 & 1749 \\
$m_{\nc}$ & 429.9 & 1025 &
& $m_{\tilde{b}_{1}}$ & 841.4 & 1729 \\
$m_{\nd}$ & 451.3 & 1029 & & $m_{\tilde{b}_{2}}$ & 970.2 & 1902 \\
$m_{\cha}$ & 208.8 & 195.5 & & $m_{\tilde{\tau}_{1}}$ & 817.7 & 1011
\\
$m_{\chb}$ & 448.6 & 1036 & & $m_{\tilde{\tau}_{2}}$ & 822.8 & 1041
\\
$m_{\tilde{g}}$ & 707.1 & 1929 & &  &  &  \\
\cline{1-7}
\end{tabular}
\end{center}}
\caption{ Relevant SUSY mass spectra for the  HWM
and PWM as calculated from the high-scale
boundary conditions given in~(\ref{5.4}) and~(\ref{5.1}), respectively.
All masses are in GeV.}\label{tbl:spectra}
\end{table}
%=================================================================

{\it Discovery Modes at the LHC:}
An important and robust aspect of supersymmetric models which are
capable of generating the observed PAMELA positron excess is that
some superpartners will necessarily be light and are therefore
potentially discoverable at the LHC. Certain key physical masses for
the HWM and PWM are given in Table~(\ref{tbl:spectra}).
All masses are computed from the high-scale boundary
conditions of Eq.(\ref{5.4}) and Eq.(\ref{5.1}) after renormalization
group evolution using SoftSUSY~\cite{ben}. One finds that for
the PWM the sum rules of Eq.(\ref{5.6}) are satisfied with an
accuracy of less than $0.5\%$. In fact, for the PWM there is an
almost perfect degeneracy between the lightest chargino and the
lightest neutralino mass, as is typical for models with a
wino-dominated LSP. In the HWM this mass difference is larger,
reflecting the larger proportion of bino and Higgsino components for
the LSP wave function, so that Eq.(\ref{5.6}) is satisfied only at
the 5-6\% level. But as we saw in Section~\ref{sec:positron} this
results in a larger $B_{\rm Co}$ factor for the HWM.

One may note that in Table~(\ref{tbl:spectra}) a 
significant difference in mass scales  for the 
$SU(3)$-charged superpartners  of the HWM and of the PWM. This difference
in the mass scales has large implications for the
discovery prospects of the two models. To analyze the signatures of
these models at the LHC we generated events using PYTHIA followed by
a detector simulation using PGS4 \cite{pythiapgs}. Two data sets for
each model at $\sqrt{s}=14$ TeV were generated for 10~fb$^{-1}$ and
100~fb$^{-1}$ of signal events, as well as a 500~pb$^{-1}$ sample at
$\sqrt{s}=10$ TeV for each model. In addition we considered a
suitably-weighted sample of 5~fb$^{-1}$ Standard Model background
events, consisting of Drell-Yan, QCD dijet, $t\,\bar{t}$,
$b\,\bar{b}$, $W$/$Z$+jets and diboson production. Events were
analyzed using level one (L1) triggers in PGS4, designed to mimic
the CMS trigger tables~\cite{Ball:2007zza}. Object-level
post-trigger cuts were also imposed. We require all photons,
electrons, muons and taus to have transverse momentum $p_T\geq 10$
GeV and $|\eta|<2.4$ and we require hadronic jets to satisfy
$|\eta|<3$. Additional post-trigger level cuts were implemented for
specific analyses, as described below.

\begin{table}[b]
    \begin{center}
\begin{tabular}{|l||c|c||c|c|}
\multicolumn{1}{c}{ } & \multicolumn{2}{c}{HWM} & \multicolumn{2}{c}{PWM} \\
 \hline Signature & Events & $S/\sqrt{B}$ & Events & $S/\sqrt{B}$ \\ \hline
Multijets & 8766 & 183.74 & 50 & 1.05 \\
Lepton + jets & 2450 & 32.25 & 26 & 0.34 \\
OS dileptons + jets & 110 & 6.39 & 4 & 0.23 \\
SS dileptons + jets & 60 & 11.77 & 0 & NA \\
Trileptons + jets & 14 & 2.47 & 0 & NA \\ \hline
\end{tabular}
{\caption{\label{tbl:discovery} LHC discovery
channels for the HWM and the PWM: Event counts are after 10~fb$^{-1}$
of integrated luminosity. All signatures require transverse
sphericity $S_T \geq 0.2$ and at least 250~GeV of $\not\!\!{E_T}$
except for the trilepton signature, where only $\not\!\!{E_T} \geq
200 \,{\rm GeV}$ is required. Here ${\rm (OS,SS) = (opposite~sign,~same~sign)}$}}
\end{center}
\end{table}

\begin{figure}[t]
  \begin{center}
\centering
 \includegraphics[width=8cm,height=7cm]{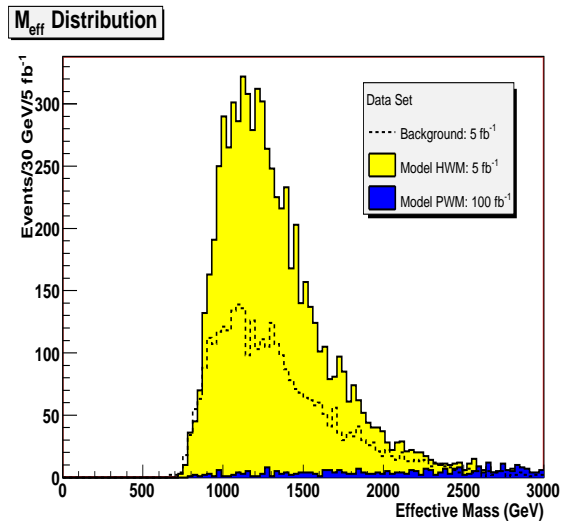}
 \caption{Effective mass distribution for the Higgsino-wino
   mixed model(HWM)  for 5~fb$^{-1}$ (yellow) and the wino model for
    100~fb$^{-1}$ (blue) along with the Standard Model background (dashed open histogram).}
    \label{graphmeff}
  \end{center} 
\end{figure}

We begin with standard SUSY discovery modes~\cite{Baer:1995nq},
slightly modified to maximize the signal significance for these
models. These five signatures are collected in
Table~(\ref{tbl:discovery}) for 10~fb$^{-1}$ of integrated luminosity.
In all cases we require transverse sphericity $S_T \geq 0.2$ and at
least 250~GeV of $\not\!\!{E_T}$ except for the trilepton signature,
where we place a cut of $\not\!\!{E_T} \geq 200 \,{\rm GeV}$. The
multijet channel includes a veto on isolated leptons and requires at
least four jets with the transverse momenta of the four leading jets
satisfying $p_T \geq \left(200,150,50,50\right)$ GeV, respectively.
For the leptonic signatures we include only $e^\pm$ and $\mu^{\pm}$
final states and demand at least two jets with the leading jets
satisfying $p_T \geq \left(100,50\right)$ GeV, respectively. For the case of the HWM, 
the existence of 
light squarks and gluinos   give rise to a relatively large number of events passing
the cuts, and thus the model  is  
easily accessible at the LHC
in almost all channels. In fact, we find that this particular model point  gives
rise to $\sim 150$ multijet events with $\not\!\!{E_T} \geq 250 \,{\rm
GeV}$ with just 500~pb$^{-1}$ at $\sqrt{s}=10$~TeV. By contrast the PWM has
nearly 2~TeV squarks and gluinos  and requires over a 100 fold increase
in luminoisty to reach a comparable event rate in this channel.  Thus after 100~fb$^{-1}$ of integrated luminosity
the PWM results in only
440~multijet events with $\not\!\!{E_T} \geq 250 \,{\rm GeV}$ . This is illustrated in
Figure~(\ref{graphmeff}) where the effective mass, defined as the
scalar sum of the transverse momenta of the four hardest jets in the
event plus the missing transverse energy, is plotted for events
satisfying the cuts described above for the multijet channel. The
signal for the HWM is clearly discernible above the Standard Model
background after 5~fb$^{-1}$. For comparison we plot the same
distribution for the PWM after 100~fb$^{-1}$.

%%%%%%%%%%%%%%%%%%%%%%%%%%%%%%%%%%%%%
%%  FIGURE 5
%%%%%%%%%%%%%%%%%%%%%%%%%%%%%%%%%%%%%

{\it Experimental Challenges of the PWM Model:}
The suppression of leptonic final states for both models, especially
the trilepton channel, is the result of the small mass difference
between the low-lying electroweak gaugino states which causes
leptonic decay products to be generally quite soft. This is
particularly severe for the wino-like scenario (the PWM). We note
that the total (leading order) supersymmetric cross section for the PWM is still a
healthy 2.3~pb (to be compared with 7.4~pb for the HWM). The lack of a 
signal here is largely the result of an inability to trigger on events in
which light electroweak gauginos are produced (99\% of the total
SUSY production cross-section) and the absence of hard leptons in
the decay products of these states. These difficulties are common to
phenomenological studies of models where the mass gaps between sparticles
are small~\cite{WinoSGURA,Chattopadhyay:2006xb,Paige:1999ui,Baer:2000bs} such as in models
with anomaly-mediated
supersymmetry breaking (AMSB)~\cite{Giudice:1998xp} which share many
of the same features in the gaugino sector~\cite{Gaillard:1999yb,Bagger:1999rd} 
as the PWM.

\begin{table}[t]
    \begin{center}
\begin{tabular}{|l||c|c||c|c|}
\multicolumn{1}{c}{ } & \multicolumn{2}{c}{HWM} & \multicolumn{2}{c}{PWM} \\
 \hline Object Cuts (GeV) & Events & $S/\sqrt{B}$ & Events & $S/\sqrt{B}$ \\ \hline
$p_T^{\rm jet} \geq 150$, $\not\!\!{E_T} \geq 150$ & 1994 & 2.13 & 3442 & 3.68 \\
$p_T^{\rm jet} \geq 200$, $\not\!\!{E_T} \geq 150$ & 1302 & 2.52 &
1983 & 3.84 \\ \hline
$p_T^{\rm jet} \geq 150$, $\not\!\!{E_T} \geq 200$ & 1334 & 2.53 & 2147 & 4.08 \\
$p_T^{\rm jet} \geq 200$, $\not\!\!{E_T} \geq 200$ & 1241 & 2.58 &
1904 & 3.95 \\ \hline
$p_T^{\rm jet} \geq 150$, $\not\!\!{E_T} \geq 300$ & 659 & 3.57 &
771 & 4.17 \\ \hline
\end{tabular}
{\caption{\label{tbl:amsb} Monojet signature for
 the HWM  and the PWM: Event counts are after 100~fb$^{-1}$ of integrated
luminosity for various choices of cuts on total $\not\!\!{E_T}$ and
jet $p_T$. All signatures involve a lepton veto and require no other
jets in the event with $p_T^{\rm jet} \geq 20$ GeV. No transverse
sphericity cut was applied in any of these signatures.}}
\end{center}
\end{table}

Nevertheless, some signatures unique to this scenario can be
explored with sufficient integrated luminosity. For example, one can
look for events which pass the initial $\not\!\!{E_T}$ trigger, but
which have no leptons or energetic jets. Such events predominantly
arise from direct production of $\cha$ and/or
${\tilde\chi}^0_{1,2}$ states whose decay products produce only
very soft objects. If the mass difference $\Delta m = m_{\cha} -
m_{\na}$ is sufficiently small the chargino will travel a
macroscopic distance before decaying. An offline analysis may then
reveal events with tracks in the inner layers of the detector which
abruptly end, leaving no further leptonic tracks or calorimeter
activity~\cite{Chen:1995yu,Feng:1999fu,Gherghetta:1999sw}. In the extreme limit
of a pure-wino LSP, as in the AMSB models, this mass difference is
typically the size of the pion mass, {\em i.e.}
$\mathcal{O}(m_{\pi})$ and the distance traveled can be several
centimeters. In the case of the PWM, however, the mass difference is
roughly twice the pion mass and the typical decay length will be
such that the decays will appear prompt and few events will reveal a
displaced vertex. More promising is to consider events with a single
high-$p_T$ jet and large missing transverse energy. Such events can
arise from initial state radiation in electroweak gaugino
production, or in cases where the lightest chargino or neutralino is
produced in association with a gluino or squark. This particular
`monojet' channel has relatively large event rates for both the HWM
and PWM, with the primary Standard Model background coming from
$W$+jets production. Event rates and signal significance for various
jet $p_T$ and $\not\!\!{E_T}$ cuts are given in
Table~(\ref{tbl:amsb}). More detailed analyses of similar models show
that such events can be bona-fide discovery modes in scenarios of
this type~\cite{Barr:2002ex}.

%%%%%%%%%%%%%%%%%%%%%%%%%%%%%%%%%%%%%
%%  FIGURE 6
%%%%%%%%%%%%%%%%%%%%%%%%%%%%%%%%%%%%%
\begin{figure}[t]
  \begin{center}
\centering
     \includegraphics[width=8cm,height=7cm]{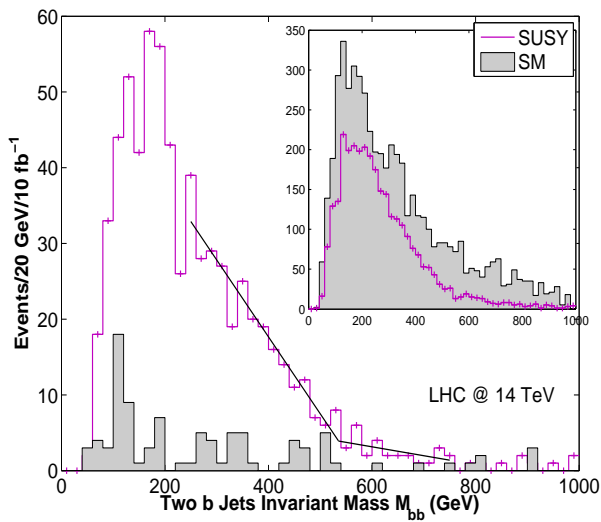}
    \caption{The invariant mass distribution of
2~b-jets events for HWM where the LSP contains substantial
Higgsino and wino components. Shaded histograms represents the
Standard Model backgrounds. The solid lines are the best fit
functions for the SUSY signal near the endpoint. The endpoint is
estimated to be $\sim$~530 GeV based on the linear fitting
functions. The embedded window shows the mass distributions for both
SUSY and SM, when one requires a 200~GeV missing energy cut. In
order to suppress the Standard Model background, we increase the
missing energy cut to 400 GeV and add two additional conditions: (1)
lepton veto; (2) at least 2 more jets besides the 2 b-tagged
jets.} \label{fig:dimass}
  \end{center}
\end{figure}

{\it Measuring Sparticle Masses in the HWM Model:}
For the mixed Higgsino-wino 
model (HWM) a sufficient number of 
events can be obtained even with a low integrated luminosity 
which will allow one to determine  
the properties of the superpartner spectrum and can confirm the features
of Table~(\ref{tbl:spectra}) -- particularly those with direct
relevance to the calculation of the thermal relic abundance and
positron yield in cosmic rays. Thus even with 10~fb$^{-1}$ it should be
possible to get a reasonable estimate of the gluino mass by
considering events with precisely two b-tagged jets. As the gluino is
relatively light it would be  produced in significant amounts and  
the invariant mass
distribution of b-jets produced in its three-body decays will reveal
a kink which allows one to determine the lower limit on the gluino
mass knowing the LSP mass~\cite{Kitano:2006gv}, {\em i.e.}
$m_{\tilde{g}} \geq \left(M_{\rm inv}^{bb}\right)^{\rm kink} +
m_{\na}$. Over 3000~events with two b-jets satisfying $p_T^{\rm jet}
\geq 60$ GeV and $\not\!\!{E_T} \geq 200$ GeV were produced in our
10~fb$^{-1}$ sample for the HWM.  With these cuts, the Standard Model
background, arising mostly from $t\,\bar{t}$ events, is comparable to the
signal as is clear from the inset of Figure~(\ref{fig:dimass}). To
reduce this background we veto events with isolated leptons, require
two additional jets without b-tags, each satisfying $p_T^{\rm jet}
\geq 60$, and increase the missing transverse energy cut from
200~GeV to 400~GeV. This reduces the signal sample by approximately
a factor of four, but the Standard Model background is now reduced
to manageable levels. This is displayed in the main panel of
Figure~(\ref{fig:dimass}) where $\left(M_{\rm inv}^{bb}\right)^{\rm
kink} \sim 500$ GeV, which gives $m_{\tilde{g}} \gappeq 700$ GeV,
which is consistent with the true gluino mass of 707~GeV in this
model.

%%%%%%%%%%%%%%%%%%%%%%%%%%%%%%%%%%%%%
%%  FIGURE 7
%%%%%%%%%%%%%%%%%%%%%%%%%%%%%%%%%%%%%
\begin{figure}[t]
  \begin{center}
\centering
 \includegraphics[width=8cm,height=7cm]{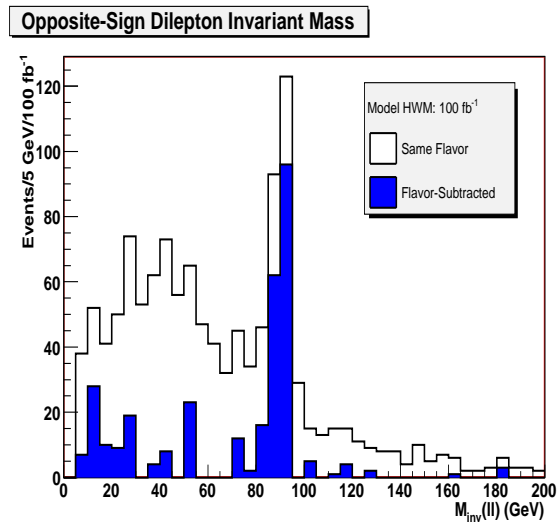}
 \caption{Dilepton invariant mass distributions for the HWM after 100~fb$^{-1}$.
 The unshaded histogram gives the invariant mass distribution for events
 with precisely two opposite-sign leptons of the same flavor (electron or muon). The
 shaded histogram gives the flavor-subtracted distribution
 which results when the opposite-sign, opposite-flavor distribution is subtracted from the
 same-flavor distribution. }\label{graphleps}
  \end{center}
\end{figure}

Of even more relevance is the relatively small mass gap between the
two lightest neutralinos $\nb$ and $\na$. While not as severe as in
the wino-like case of the PWM, the mass difference is still small
enough that the number of energetic leptons coming from the final
stages of sparticle cascade decays will be suppressed relative to
the number expected from more universal models. The reduction in
events with two or more energetic, isolated leptons will
significantly degrade the ability to make measurements of mass
differences using the edges of various kinematic distributions. For
example, a typical strategy for accessing the mass difference
between light neutralinos is to form the flavor-subtracted dilepton
invariant mass for events with at least two jets satisfying
$p_T^{\rm jet} \geq 60$~GeV, at least 200~GeV of $\not\!\!{E_T}$ and
two opposite-sign leptons~\cite{Hinchliffe:1996iu}. The invariant
mass distribution formed from the subset involving two leptons of
opposite flavor is subtracted from that involving two of the same
flavor, i.e., the combination ($e^+e^-+\mu^+\mu^--e^+\mu^--e^-\mu^+$), to reduce background. In Fig.(\ref{graphleps}) 
we plot the
invariant mass of same-flavor, opposite-sign leptons in two-lepton
events for HWM after 100~fb$^{-1}$ as well as the
flavor-subtracted distribution. Note the small number of events
which remain after the subtraction procedure has been performed.
Nevertheless, the beginnings of a feature in the low-energy bins can
be discerned which is consistent with the mass differences between
the low-lying gauginos in this model. Additional statistics and more
careful analysis techniques will be needed to make strong statements
about the neutralino masses in this model.

%%%%%%%%%%%%%%%%%%%%%%%%%%%%%%%%%%%%%
{\it ILC Implications:}
The ideal machine  
to study the spectroscopy of the light gauginos
and to confirm the model as a potential explanation for the PAMELA
positron excess would be at a potential International Linear
Collider (ILC). We emphasize that a linear collider operating at a
center-of-mass energy of 500~GeV  would
be sufficient to study the closely-spaced lightest neutralinos and
lightest charginos of either model. For the case of the pure wino model (PWM) a
future linear collider will be essential for resolving the presence
of two nearly-degenerate states near 200~GeV and for studying their
couplings. The prospects for both models at the ILC are quite good
with $\sigma(e^+e^-\to {\rm SUSY}) = O(0.1)\,{\rm pb}$ for both
models. Finally, it is worth pointing out that for both models there
is the distinct possibility of observing a degenerate cluster of
$Z'$ bosons with masses of the order of the lightest chargino mass.
This will only be possible if the decays of such additional $Z'$
bosons into hidden sector matter states are forbidden or largely
suppressed. In such cases they would appear as sharp resonances in
Drell-Yan processes
and should be discernible at the LHC with
sufficient  luminosity. Observation of such states would be a spectacular confirmation of
the $B_{\rm Co}$ mechanism of generating the correct neutralino
relic density by thermal means for the HWM model.

\noindent
\vspace{-.6cm}
\section{Conclusions}
\vspace{-.5cm}
The positron excess observed in the PAMELA satellite experiment has
spawned various mechanisms to explain this effect. An interesting
possibility relates to the positron excess arising from the
annihilation of dark matter in the galaxy. An adequate explanation
of this phenomenon within a particle physics model then would
require a simultaneous fit both to the WMAP data regarding the
density of cold dark matter as well as to the PAMELA data. Often it
turns  out that models that give the desired relic density give too
small a $\langle \sigma v \rangle$ in the galaxy to explain the
PAMELA excess. Alternately, models which produce an adequate
$\langle \sigma v \rangle$ and explain the PAMELA excess fail to
produce  the proper relic density.  To reconcile the two phenomena
typically the following mechanisms have been proposed: (i) For
models that generate the right relic density but give too small a
$\langle \sigma v \rangle$ the Sommerfeld enhancement or the
Breit-Wigner pole enhancement
of $ \langle
\sigma v \rangle $ can explain both sets of data; (ii) For models
which give an adequate  $\langle \sigma v \rangle$ to explain the
PAMELA excess but give too small a relic density, a non-thermal
mechanism
 is taken to produce the proper relic
abundance.

In this paper we have carried out an analysis of a distinctly different mechanism
which can generate the proper relic density for the second type of
models, {\em i.e.}, models where  $\langle \sigma v \rangle$ is
large enough to explain the PAMELA excess but the relic density
needs an enhancement. We illustrate this mechanism within the
framework of nonuniversal SUGRA  models with  an
extended hidden $U(1)^n_X$ gauge symmetry. The extension along with
conditions of mass degeneracy in the hidden sector gives rise to
predictions which match the PAMELA data and the predicted relic
density is consistent with the WMAP data with a mixed Higgsino-wino
LSP. The anti-proton and the gamma ray fluxes emanating from the
annihilation of dark matter in the galaxy are found to be compatible
with data. Implications of the model for the direct detection of
dark matter and some of its collider signatures were also discussed.
The model is testable on both fronts. Specifically the mixed
Higgsino-wino LSP model is testable at the LHC with just 1~fb$^{-1}$
of data. While the analysis was done in the framework of
nonuniversal SUGRA models, the results of the analysis are valid in a
broad class of models including string and D-brane models as long
there is an LSP with mass in the proper range and a  suitable admixture of
Higgsino-wino components 
with an appropriate degeneracy in the
hidden sector. Finally, we note that we have not made an
attempt here to fit the FERMI-LAT $e^{+} + e^-$
result~\cite{Abdo:2009zk}. Such a fit can be accommodated by
assuming that the high energy flux is a consequence of pulsars or
mixed dark matter and pulsar contributions~\cite{Yuksel,fermilat}. 
\\
\vspace{-.025cm}
\noindent
{\em Acknowledgments}:
This research is supported in part by NSF Grants
No. PHY-0757959 and No. PHY-0653587 (Northeastern)
and No. PHY-0653342 (Stony Brook), and by the US Department 
of Energy (DOE) under grant DE-FG02-95ER40899 (Michigan).

%%%%%%%%%%%%%%%%%%%%%%%%%%%%


\begin{thebibliography}{999}
\bibitem{pamela}
  O.~Adriani {\it et al.}  [PAMELA Collaboration],
  %``An anomalous positron abundance in cosmic rays with energies 1.5.100 GeV,''
  Nature {\bf 458}, 607 (2009);
 % [arXiv:0810.4995 [astro-ph]].
    %%CITATION = PRLTA,102,051101;%%
  %``A new measurement of the antiproton-to-proton flux ratio up to 100 GeV in
  %the cosmic radiation,''
  Phys.\ Rev.\ Lett.\  {\bf 102}, 051101 (2009).
%  [arXiv:0810.4994 [astro-ph]].
  %%CITATION = PRLTA,102,051101;%%

\bibitem{Abdo:2009zk}
  A.~A.~Abdo {\it et al.}  [The Fermi LAT Collaboration],
  %``Measurement of the Cosmic Ray e+ plus e- spectrum from 20 GeV to 1 TeV with
  %the Fermi Large Area Telescope,''
  Phys.\ Rev.\ Lett.\  {\bf 102}, 181101 (2009);
%  [arXiv:0905.0025 [astro-ph.HE]];
  %%CITATION = PRLTA,102,181101;%%
  See also,
 J.~Chang {\it et al.},
  %``An Excess Of Cosmic Ray Electrons At Energies Of 300.800 Gev,''
  Nature {\bf 456}, 362 (2008).
  %%CITATION = NATUA,456,362;%%

\bibitem{Cirelli:2008id}
  M.~Cirelli, R.~Franceschini and A.~Strumia,
  %``Minimal Dark Matter predictions for galactic positrons, anti-protons,
  %photons,''
  Nucl.\ Phys.\  B {\bf 800}, 204 (2008);
  M.~Cirelli, M.~Kadastik, M.~Raidal, A.~Strumia,
  %``Model-independent implications of the e+, e-, anti-proton cosmic ray
  %spectra on properties of Dark Matter,''
  Nucl.\ Phys.\  B {\bf 813}, 1 (2009).
  %[arXiv:0802.3378 [hep-ph]].
  %%CITATION = NUPHA,B800,204;%%

 \bibitem{kaneetal}
  P.~Grajek, G.~Kane, D.~J.~Phalen, A.~Pierce and S.~Watson,
  %``Neutralino Dark Matter from Indirect Detection Revisited,''
  arXiv:0807.1508 [hep-ph];
  %%CITATION = ARXIV:0807.1508;%
  %``Is the PAMELA Positron Excess Winos?,''
 Phys.\ Rev.\  D {\bf 79}, 043506   (2009);
% [ arXiv:0812.4555 [hep-ph]];
  %%CITATION = ARXIV:0812.4555;%%
    G.~Kane, R.~Lu and S.~Watson,
  %``PAMELA Satellite Data as a Signal of Non-Thermal Wino LSP Dark Matter,''
  arXiv:0906.4765 [astro-ph.HE].
  %%CITATION = ARXIV:0906.4765;%%

 \bibitem{Hisano08}
  J.~Hisano, M.~Kawasaki, K.~Kohri and K.~Nakayama,
  %``Positron/Gamma-Ray Signatures of Dark Matter Annihilation and Big-Bang
  %Nucleosynthesis,''
  Phys.\ Rev.\  D {\bf 79}, 063514 (2009).
 % [arXiv:0810.1892 [hep-ph]].
  %%CITATION = PHRVA,D79,063514;%%

\bibitem{Feldman:2008xs}
  D.~Feldman, Z.~Liu and P.~Nath,
  %``PAMELA Positron Excess as a Signal from the Hidden Sector,''
  Phys.\ Rev.\  D {\bf 79}, 063509 (2009);
%  [arXiv: 0810.5762 [hep-ph]];
  %%CITATION = PHRVA,D79,063509;%%
  M.~Ibe, H.~Murayama and T.~T.~Yanagida,
  %``Breit-Wigner Enhancement of Dark Matter Annihilation,''
  Phys.\ Rev.\  D {\bf 79}, 095009 (2009);
  W.~L.~Guo and Y.~L.~Wu,
  %``Enhancement of Dark Matter Annihilation via Breit-Wigner Resonance,''
  Phys.\ Rev.\  D {\bf 79}, 055012 (2009);
  W.~L.~Guo and X.~Zhang,
  %``Constraints on Dark Matter Annihilation Cross Section in Scenarios of
  %Brane-World and Quintessence,''
  arXiv:0904.2451 [hep-ph].
  M.~Ibe, Y.~Nakayama, H.~Murayama and T.~T.~Yanagida,
  %``Nambu-Goldstone Dark Matter and Cosmic Ray Electron and Positron Excess,''
  arXiv:0902.2914 [hep-ph];
    X.~J.~Bi, X.~G.~He and Q.~Yuan,
  %``Parameters in a Class of Leptophilic Dark Matter Models from ATIC and
  %PAMELA,''
  arXiv:0903.0122 [hep-ph];
  %%CITATION = ARXIV:0903.0122;%%
  F.~Y.~Cyr-Racine, S.~Profumo and K.~Sigurdson,
  %``Protohalo Constraints to the Resonant Annihilation of Dark Matter,''
  arXiv:0904.3933 [astro-ph.CO];
 For related work see: 
D.~Feldman, Z.~Liu and P.~Nath,
  %``The Stueckelberg Z' extension with kinetic mixing and milli-charged dark
  %matter from the hidden sector,''
  Phys.\ Rev.\  D {\bf 75}, 115001 (2007);
    M.~Pospelov and A.~Ritz,
  %``Astrophysical Signatures of Secluded Dark Matter,''
  Phys.\ Lett.\  B {\bf 671}, 391 (2009); W.~Shepherd, T.~M.~P.~Tait and G.~Zaharijas,
Phys.\  Rev.\  D {\bf 79}, 055022 (2009).



\bibitem{barger}
  V.~Barger, W.~Y.~Keung, D.~Marfatia and G.~Shaughnessy,
  %``PAMELA and dark matter,''
  Phys.\ Lett.\  B {\bf 672}, 141 (2009);
    Y.~Nomura and J.~Thaler,
  %``Dark Matter through the Axion Portal,''
  arXiv:0810.5397 [hep-ph];
  %[arXiv:0809.0162 [hep-ph]];
  J.~Zhang, X.~J.~Bi, J.~Liu, S.~M.~Liu, P.~f.~Yin, Q.~Yuan and S.~H.~Zhu,
  %``Discriminating different scenarios to account for the PAMELA and ATIC data
  %by synchrotron and IC radiation,''
  arXiv:0812.0522 [astro-ph];
  %%CITATION = ARXIV:0812.0522;%%
  %%CITATION = JCAPA,0903,009;%%
 % [arXiv:0809.2409 [hep-ph]].
  %%CITATION = NUPHA,B813,1;%%
  %%CITATION = PHLTA,B671,391;%%
  P.~H.~Frampton and P.~Q.~Hung,
  %``Positron Excess, Luminous-Dark Matter Unification and Family Structure,''
  Phys.\ Lett.\  B {\bf 675}, 411 (2009);
  S.~L.~Chen, R.~N.~Mohapatra, S.~Nussinov and Y.~Zhang,
  %``R-Parity Breaking via Type II Seesaw, Decaying Gravitino Dark Matter and
  %PAMELA Positron Excess,''
  Phys.\ Lett.\  B {\bf 677}, 311 (2009);
  %%CITATION = PHLTA,B677,311;%%
  %%CITATION = PHLTA,B675,411;%%
%\cite{Cheung:2009si}
  K.~Cheung, P.~Y.~Tseng and T.~C.~Yuan,
  %``Double-action dark matter, PAMELA and ATIC,''
  arXiv:0902.4035 [hep-ph];
  %%CITATION = ARXIV:0902.4035;%%
  I.~Gogoladze, N.~Okada and Q.~Shafi,
  %``Type II Seesaw and the PAMELA/ATIC Signals,''
  arXiv:0904.2201 [hep-ph];
  %%CITATION = ARXIV:0904.2201;%%
   J.~Hisano, M.~Kawasaki, K.~Kohri, T.~Moroi and K.~Nakayama,
  %``Cosmic Rays from Dark Matter Annihilation and Big-Bang Nucleosynthesis,''
  Phys.\ Rev.\  D {\bf 79}, 083522 (2009);
  %%CITATION = PHRVA,D79,083522;%%
  %%CITATION = PHLTA,B672,141;%%
  F.~Chen, J.~M.~Cline and A.~R.~Frey,
  %``Nonabelian dark matter: models and constraints,''
  arXiv:0907.4746 [hep-ph].
  %%CITATION = ARXIV:0907.4746;%%

\bibitem{other}
  H.~Davoudiasl,
  %``Dark Matter with Time-Varying Leptophilic Couplings,''
  arXiv:0904.3103 [hep-ph];
  %%CITATION = ARXIV:0904.3103;%
  %%CITATION = ASJOA,699,L59;%%
%\cite{Cao:2009yy}
%\cite{Brandenberger:2009ia}
  R.~Brandenberger, Y.~F.~Cai, W.~Xue and X.~m.~Zhang,
  %``Cosmic Ray Positrons from Cosmic Strings,''
  arXiv:0901.3474 [hep-ph];
  %%CITATION = ARXIV:0901.3474;%%
  Q.~H.~Cao, E.~Ma and G.~Shaughnessy,
  %``Dark Matter: The Leptonic Connection,''
  Phys.\ Lett.\  B {\bf 673}, 152 (2009);
  %%CITATION = PHLTA,B673,152;%%
  C.~R.~Chen, M.~M.~Nojiri, S.~C.~Park, J.~Shu and M.~Takeuchi,
  %``Dark matter and collider phenomenology of split-UED,''
  arXiv:0903.1971 [hep-ph];
%
  S.~C.~Park and J.~Shu,
  %``Split-UED and Dark Matter,''
  Phys.\ Rev.\  D {\bf 79}, 091702 (2009);
 % [arXiv:0901.0720 [hep-ph]].
  %%CITATION = PHRVA,D79,091702;%%
  %%CITATION = ARXIV:0903.1971;%%
  %%CITATION = ARXIV:0903.3625;%%
%\cite{Bergstrom:2009ib}
  L.~Bergstrom,
  %``Dark Matter Candidates,''
  arXiv:0903.4849 [hep-ph];
  %%CITATION = ARXIV:0903.4849;%%
%\cite{Braeuninger:2009pe}
  C.~B.~Braeuninger and M.~Cirelli,
  %``Anti-deuterons from heavy Dark Matter,''
  Phys.\ Lett.\  B {\bf 678}, 20 (2009);
  %[arXiv:0904.1165 [hep-ph]]
  %%CITATION = PHLTA,B678,20;%%
  Q.~Yuan, X.~J.~Bi, J.~Liu, P.~F.~Yin, J.~Zhang and S.~H.~Zhu,
  %``Clumpiness enhancement of charged cosmic rays from dark matter annihilation
  %with Sommerfeld effect,''
  arXiv:0905.2736 [astro-ph.HE];
  %%CITATION = ARXIV:0905.2736;%%
  N.~Okada and T.~Yamada,
  %``The PAMELA and Fermi signals from long-lived Kaluza-Klein dark matter,''
  arXiv:0905.2801 [hep-ph];
  %%CITATION = ARXIV:0905.2801;%%
  C.~H.~Chen,
  %``Resolution to neutrino masses, baryon asymmetry in leptogenesis and
  %cosmic-ray anomalies,''
  arXiv:0905.3425 [hep-ph];
  %%CITATION = ARXIV:0905.3425;%%
  P.~H.~Gu, H.~J.~He, U.~Sarkar and X.~m.~Zhang,
  %``Double Type-II Seesaw, Baryon Asymmetry and Dark Matter for Cosmic e^\pm
  %Excesses,''
  arXiv:0906.0442 [hep-ph];
  %%CITATION = ARXIV:0906.0442;%%
   K.~Kohri, J.~McDonald and N.~Sahu,
  %``Cosmic Ray Anomalies and Dark Matter Annihilation to Muons via a Higgs
  %Portal Hidden Sector,''
  arXiv:0905.1312 [hep-ph];
  %%CITATION = ARXIV:0905.1312;%%
  C.~Balazs, N.~Sahu and A.~Mazumdar,
  %``Absolute electron and positron fluxes from PAMELA/Fermi and Dark Matter,''
  arXiv:0905.4302 [hep-ph].
  %%CITATION = ARXIV:0905.4302;%%

\bibitem{decay}
 A.~Ibarra, A.~Ringwald, D.~Tran and C.~Weniger,
  %``Cosmic Rays from Leptophilic Dark Matter Decay via Kinetic Mixing,''
  arXiv:0903.3625 [hep-ph];
  C.~R.~Chen, M.~M.~Nojiri, F.~Takahashi and T.~T.~Yanagida,
  %``Decaying Hidden Gauge Boson and the PAMELA and ATIC/PPB-BETS Anomalies,''
  arXiv:0811.3357 [astro-ph];
  %%CITATION = ARXIV:0811.3357;%%
  E.~Nardi, F.~Sannino and A.~Strumia,
  %``Decaying Dark Matter can explain the electron/positron excesses,''
  JCAP {\bf 0901}, 043 (2009);
 % [arXiv:0811.4153 [hep-ph]];
  J.~Liu, P.~f.~Yin and S.~h.~Zhu,
  %``Prospects for Detecting Neutrino Signals from Annihilating/Decaying Dark
  %Matter to Account for the PAMELA and ATIC results,''
  arXiv:0812.0964 [astro-ph];
  %%CITATION = ARXIV:0812.0964;%%
  %%CITATION = JCAPA,0901,043;%
 C.~R.~Chen, K.~Hamaguchi, M.~M.~Nojiri, F.~Takahashi and S.~Torii,
  %``Dark Matter Model Selection and the ATIC/PPB-BETS anomaly,''
  JCAP {\bf 0905}, 015 (2009);
  %[arXiv:0812.4200 [astro-ph]];
  %%CITATION = JCAPA,0905,015;%%
%\cite{Ishiwata:2008qy}
  K.~Ishiwata, S.~Matsumoto and T.~Moroi,
  %``Synchrotron Radiation from the Galactic Center in Decaying Dark Matter
  %Scenario,''
  Phys.\ Rev.\  D {\bf 79}, 043527 (2009);
  %``High Energy Cosmic Rays from Decaying Supersymmetric Dark Matter,''}
  %``Cosmic Gamma-ray from Inverse Compton Process in Unstable Dark Matter
  %Scenario,''
 % arXiv:0905.4593 [astro-ph.CO];
  %%CITATION = ARXIV:0905.4593;%%
  JHEP {\bf 0905}, 110 (2009);
  %%CITATION = JHEPA,0905,110;%%
 %%CITATION = PHRVA,D79,043527;%%
  X.~Chen,
  %``Decaying Hidden Dark Matter in Warped Compactification,''
  arXiv:0902.0008 [hep-ph];
  %%CITATION = ARXIV:0902.0008;%
  H.~Fukuoka, J.~Kubo and D.~Suematsu,
  %``Anomaly Induced Dark Matter Decay and PAMELA/ATIC Experiments,''
  Phys.\ Lett.\  B {\bf 678}, 401 (2009).
  %%CITATION = PHLTA,B678,401;%%
 

\bibitem{susyattempts}
  L.~Bergstrom, T.~Bringmann and J.~Edsjo,
  %``New Positron Spectral Features from Supersymmetric Dark Matter - a Way to
  %Explain the PAMELA Data?,''
  Phys.\ Rev.\  D {\bf 78}, 103520 (2008);
     R.~Harnik and G.~D.~Kribs,
  %``An Effective Theory of Dirac Dark Matter,''
  arXiv:0810.5557 [hep-ph];
  %%CITATION = ARXIV:0810.5557;%%
  %[arXiv:0808.3725 [astro-ph]].
  %%CITATION = PHRVA,D78,103520;%%
  R.~Allahverdi, B.~Dutta, K.~Richardson-McDaniel and Y.~Santoso,
  %``A Supersymmetric $B^-$ L Dark Matter Model and the Observed Anomalies in
  %the Cosmic Rays,''
  Phys.\ Rev.\  D {\bf 79}, 075005 (2009);
%  [arXiv:0812.2196 [hep-ph]];
  %%CITATION = PHRVA,D79,075005;%%
%  R.~Allahverdi, B.~Dutta, K.~Richardson-McDaniel and Y.~Santoso,
  %``Sneutrino Dark Matter and the Observed Anomalies in Cosmic Rays,''
  Phys.\ Lett.\  B {\bf 677}, 172 (2009);
  %%CITATION = PHLTA,B677,172;%%
%\cite{Goh:2009wg}
  H.~S.~Goh, L.~J.~Hall and P.~Kumar,
  %``The Leptonic Higgs as a Messenger of Dark Matter,''
  JHEP {\bf 0905}, 097 (2009);
  %%CITATION = JHEPA,0905,097;%%
  %\cite{Chen:2009mj}
  C.~H.~Chen, C.~Q.~Geng and D.~V.~Zhuridov,
  %``Resolving Fermi, PAMELA and ATIC anomalies in split supersymmetry without
  %R-parity,''
  arXiv:0905.0652 [hep-ph];
  Y.~Bai, M.~Carena and J.~Lykken,
  %``The PAMELA excess from neutralino annihilation in the NMSSM,''
  arXiv:0905.2964 [hep-ph];
  %%CITATION = ARXIV:0905.2964;%
  %%CITATION = ARXIV:0905.0652;%%
  D.~A.~Demir, L.~L.~Everett, M.~Frank, L.~Selbuz and I.~Turan,
  %``Sneutrino Dark Matter: Symmetry Protection and Cosmic Ray Anomalies,''
  arXiv:0906.3540 [hep-ph].
  %%CITATION = ARXIV:0906.3540;%%

\bibitem{photon}
  G.~Bertone, M.~Cirelli, A.~Strumia and M.~Taoso,
  %``Gamma-ray and radio tests of the e+e- excess from DM annihilations,''
  JCAP {\bf 0903}, 009 (2009);
  L.~Bergstrom, G.~Bertone, T.~Bringmann, J.~Edsjo and M.~Taoso,
  %``Gamma-ray and Radio Constraints of High Positron Rate Dark Matter Models
  %Annihilating into New Light Particles,''
  Phys.\ Rev.\  D {\bf 79}, 081303 (2009);
  %%CITATION = PHRVA,D79,081303;%%
  W.~de Boer,
  %``Indirect Dark Matter Signals from EGRET and PAMELA compared,''
  arXiv:0901.2941 [hep-ph];
  %%CITATION = ARXIV:0901.2941;%
  %\cite{Kawasaki:2009nr}
  M.~Kawasaki, K.~Kohri and K.~Nakayama,
  %``Diffuse gamma-ray background and cosmic-ray positrons from annihilating
  %dark matter,''
  arXiv:0904.3626 [astro-ph.CO];
  %%CITATION = ARXIV:0904.3626;%%
  %%CITATION = ARXIV:0905.0480;%%
  %%CITATION = ARXIV:0901.2925;%%
  X.~J.~Bi, R.~Brandenberger, P.~Gondolo, T.~Li, Q.~Yuan and X.~Zhang,
  %``Non-Thermal Production of WIMPs, Cosmic $e^\pm$ Excesses and $\gamma$-rays
  %from the Galactic Center,''
  arXiv:0905.1253 [hep-ph];
  %%CITATION = ARXIV:0905.1253;%%
  R.~Essig, N.~Sehgal and L.~E.~Strigari,
  %``Bounds on Cross-sections and Lifetimes for Dark Matter Annihilation and
  %Decay into Charged Leptons from Gamma-ray Observations of Dwarf Galaxies,''
  Phys.\ Rev.\  D {\bf 80}, 023506 (2009).
  %%CITATION = PHRVA,D80,023506;%%

\bibitem{neutrinos}
%\cite{Delaunay:2008pc}
  C.~Delaunay, P.~J.~Fox and G.~Perez,
  %``Probing Dark Matter Dynamics via Earthborn Neutrinos at IceCube,''
  JHEP {\bf 0905}, 099 (2009);
  %%CITATION = JHEPA,0905,099;%%
  J.~Hisano, K.~Nakayama and M.~J.~S.~Yang,
  %``Upward muon signals at neutrino detectors as a probe of dark matter
  %properties,''
  Phys.\ Lett.\  B {\bf 678}, 101 (2009);
  %%CITATION = PHLTA,B678,101;%%
   M.~R.~Buckley, K.~Freese, D.~Hooper, D.~Spolyar and H.~Murayama,
  %``High-Energy Neutrino Signatures of Dark Matter Decaying into Leptons,''
  arXiv:0907.2385 [astro-ph.HE].
  %%CITATION = ARXIV:0907.2385;%%








\bibitem{Yuksel}
  H.~Yuksel, M.~D.~Kistler and T.~Stanev,
  %``TeV Gamma Rays from Geminga and the Origin of the GeV Positron Excess,''
  %arXiv:0810.2784 [astro-ph];
  Phys.\ Rev.\ Lett.\  {\bf 103}  051101 (2009);
  %%CITATION = ARXIV:0810.2784;%%
  D.~Hooper, P.~Blasi and P.~D.~Serpico,
  %``Pulsars as the Sources of High Energy Cosmic Ray Positrons,''
  JCAP {\bf 0901}, 025 (2009);
    S.~Profumo,
  %``Dissecting Pamela (and ATIC) with Occam's Razor: existing, well-known
  %Pulsars naturally account for the 'anomalous' Cosmic-Ray Electron and
  %Positron Data,''
  arXiv:0812.4457 [astro-ph];
  %%CITATION = ARXIV:0812.4457;%%
 D.~Hooper, A.~Stebbins and K.~M.~Zurek,
 %``The PAMELA and ATIC Excesses From a Nearby Clump of Neutralino Dark
 %Matter,''
 arXiv:0812.3202 [hep-ph];
 %%CITATION = ARXIV:0812.3202;%%
  E.~Borriello, A.~Cuoco and G.~Miele,
  %``Secondary radiation from the Pamela/ATIC excess and relevance for Fermi,''
  Astrophys.\ J.\  {\bf 699}, L59 (2009).
  V.~Barger, Y.~Gao, W.~Y.~Keung, D.~Marfatia and G.~Shaughnessy,
  %``Dark matter and pulsar signals for Fermi LAT, PAMELA, ATIC, HESS and WMAP
  %data,''
  arXiv:0904.2001 [hep-ph]; See also the last Ref. of \cite{kaneetal}.
  %%CITATION = ARXIV:0904.2001;%%

\bibitem{msugra}
A.~H.~Chamseddine, R.~Arnowitt and P.~Nath,
  %``Locally Supersymmetric Grand Unification,''
  Phys.\ Rev.\ Lett.\  {\bf 49} (1982) 970;
  %%CITATION = PRLTA,49,970;%%
  P.~Nath, R.~L.~Arnowitt and A.~H.~Chamseddine,
  %``Gauge Hierarchy In Supergravity Guts,''
  Nucl.\ Phys.\  B {\bf 227}, 121 (1983);
  %%CITATION = NUPHA,B227,121;%%
 L. Hall, J. Lykken and S. Weinberg, Phys. Rev. {\bf D27}, 2359 (1983);
 For reviews see:
 P.~Nath, R.~L.~Arnowitt and A.~H.~Chamseddine,
  ``Applied N=1 Supergravity,'', Trieste Particle Phys.1983:1 (QCD161:W626:1983),
World Scientific (July 1984);
  %%CITATION = NUB-2613;%%
 H.~P.~Nilles,
  %``Supersymmetry, Supergravity And Particle Physics,''
  Phys.\ Rept.\  {\bf 110}, 1 (1984);
  %%CITATION = PRPLC,110,1;%%
P.~Nath,
%  ``Twenty years of SUGRA,''
  arXiv:hep-ph/0307123.
  %%CITATION = HEP-PH/0307123;%%
%\cite{Haber:1984rc}

\bibitem{Turner:1989kg}
  M.~S.~Turner and F.~Wilczek,
 %  ``Positron Line Radiation from Halo WIMP Annihilations as a Dark Matter Signature,''
  Phys.\ Rev.\  D {\bf 42}, 1001 (1990).
  %%CITATION = PHRVA,D42,1001;%%
%\cite{Griest:1989zh}

\bibitem{WMAP}
  D.~N.~Spergel {\it et al.}  [WMAP Collaboration],
  %``Wilkinson Microwave Anisotropy Probe (WMAP) three year results:
  %Implications for cosmology,''
  Astrophys.\ J.\ Suppl.\  {\bf 170}, 377 (2007);
  E.~Komatsu {\it et al.}  [WMAP Collaboration],
  %``Five-Year Wilkinson Microwave Anisotropy Probe (WMAP\altaffilmark 1 )
  %Observations:Cosmological Interpretation,''
  Astrophys.\ J.\ Suppl.\  {\bf 180}, 330 (2009).
  %%CITATION = APJSA,170,377;%%

\bibitem{Moroi:1999zb}
  T.~Moroi and L.~Randall,
  %``Wino cold dark matter from anomaly-mediated SUSY breaking,''
  Nucl.\ Phys.\  B {\bf 570}, 455 (2000);
%  [arXiv:hep-ph/9906527].
  %%CITATION = NUPHA,B570,455;%%
  G.~L.~Kane, L.~T.~Wang and J.~D.~Wells,
  %``Supersymmetry and the positron excess in cosmic rays,''
  Phys.\ Rev.\  D {\bf 65}, 057701 (2002);
%  [arXiv:hep-ph/0108138].
  %%CITATION = PHRVA,D65,057701;%%
  B.~S.~Acharya, P.~Kumar, K.~Bobkov, G.~Kane, J.~Shao and S.~Watson,
  %``Non-thermal Dark Matter and the Moduli Problem in String Frameworks,''
  JHEP {\bf 0806}, 064 (2008).
%  [arXiv:0804.0863 [hep-ph]].
  %%CITATION = JHEPA,0806,064;%%

\bibitem{NU}
 J.R.~Ellis, K.~Enqvist, D.V.~Nanopoulos and K.~Tamvakis,
  %``Gaugino Masses And Grand Unification,''
  Phys.\ Lett.\  B {\bf 155}, 381 (1985);
  %%CITATION = PHLTA,B155,381;%%
  M.~Drees,
  %``Phenomenological Consequences Of N=1 Supergravity Theories With Nonminimal
  %Kinetic Energy Terms For Vector Superfields,''
  Phys.\ Lett.\  B {\bf 158}, 409 (1985).
  %%CITATION = PHLTA,B158,409;%%

\bibitem{NUSGURA}
   A.~Corsetti and P.~Nath,
  %``Gaugino Mass Nonuniversality and Dark Matter in SUGRA, Strings and D Brane
  %Models,''
  Phys.\ Rev.\  D {\bf 64}, 125010 (2001);
%  [arXiv:hep-ph/0003186];
%  [arXiv:hep-ph/0102075].
  %%CITATION = PHRVA,D64,015008;%%
  %%CITATION = PHRVA,D64,125010;%%
 U.~Chattopadhyay and P.~Nath,
  %``b - tau unification, g(mu)-2, the b --> s + gamma constraint and
  %nonuniversalities,''
  Phys.\ Rev.\  D {\bf 65}, 075009 (2002);
%  [arXiv:hep-ph/0110341];
  %%CITATION = PHRVA,D65,075009;%%
    G.~L.~Kane, J.~D.~Lykken, S.~Mrenna, B.~D.~Nelson, L.~T.~Wang and T.~T.~Wang,
  %``Theory-motivated benchmark models and superpartners at the Tevatron,''
  Phys.\ Rev.\  D {\bf 67}, 045008 (2003);
%  [arXiv:hep-ph/0209061].
  %%CITATION = PHRVA,D67,045008;%%
  A.~Birkedal-Hansen and B.~D.~Nelson,
  %``Relic neutralino densities and detection rates with nonuniversal gaugino
  %masses,''
  Phys.\ Rev.\  D {\bf 67}, 095006 (2003);
  D.~G.~Cerdeno and C.~Munoz,
  %``Neutralino dark matter in supergravity theories with non-universal  scalar
  %and gaugino masses,''
  JHEP {\bf 0410}, 015 (2004);
  %[arXiv:hep-ph/0405057].
  %%CITATION = JHEPA,0410,015;%%
    H.~Baer, A.~Mustafayev, S.~Profumo, A.~Belyaev and X.~Tata,
  %``Direct, indirect and collider detection of neutralino dark matter in SUSY
  %models with non-universal Higgs masses,''
  JHEP {\bf 0507}, 065 (2005);
   S.~F.~King, J.~P.~Roberts and D.~P.~Roy,
  %``Natural Dark Matter in SUSY GUTs with Non-universal Gaugino Masses,''
  JHEP {\bf 0710}, 106 (2007).
%  [arXiv:0705.4219 [hep-ph]].

\bibitem{WinoSGURA}
  A.~Birkedal-Hansen and B.~D.~Nelson,
  %``The role of Wino content in neutralino dark matter,''
  Phys.\ Rev.\  D {\bf 64}, 015008 (2001).

  \bibitem{NUSGURAColliders}
   D.~Feldman, Z.~Liu and P.~Nath,
  %``Light Higgses at the Tevatron and at the LHC and Observable Dark Matter in
  %SUGRA and D Branes,''
  Phys.\ Lett.\  B {\bf 662}, 190 (2008);
  %[arXiv:0711.4591 [hep-ph]].
  %%CITATION = PHLTA,B662,190;%%
    H.~Baer, A.~Mustafayev, E.~K.~Park and X.~Tata,
  %``Collider signals and neutralino dark matter detection in
  %relic-density-consistent models without universality,''
  JHEP {\bf 0805}, 058 (2008);
 D.~Feldman, Z.~Liu and P.~Nath,
  %``Sparticles at the LHC,''
  JHEP {\bf 0804}, 054 (2008);
   B.~Altunkaynak, M.~Holmes and B.~D.~Nelson,
  %``Solving the LHC Inverse Problem with Dark Matter Observations,''
  JHEP {\bf 0810}, 013 (2008);
  %[arXiv:0804.2899 [hep-ph]].
   S.~P.~Martin,
  %``Exploring compressed supersymmetry with same-sign top quarks at the Large
  %Hadron Collider,''
  Phys.\ Rev.\  D {\bf 78} (2008) 055019;
  %[arXiv:0807.2820 [hep-ph]].
      S.~Bhattacharya, A.~Datta and B.~Mukhopadhyaya,
  %``Non-universal gaugino and scalar masses, hadronically quiet trileptons and
  %the Large Hadron Collider,''
  Phys.\ Rev.\  D {\bf 78}, 115018 (2008);
    U.~Chattopadhyay and D.~Das,
  %``Higgs funnel region of SUSY dark matter for small tan beta, RG effects on
  %pseudoscalar Higgs boson with scalar mass non-universality,''
  Phys.\ Rev.\  D {\bf 79}, 035007 (2009);
  %[arXiv:0809.4065 [hep-ph]].
  %%CITATION = PHRVA,D79,035007;%%
  %[arXiv:0802.4085 [hep-ph]];
  %%CITATION = JHEPA,0804,054;%%
 B.~Altunkaynak, P.~Grajek, M.~Holmes, G.~Kane and B.~D.~Nelson,
  %``Studying Gaugino Mass Unification at the LHC,''
  JHEP {\bf 0904}, 114 (2009);
  %[arXiv:0901.4013 [hep-ph]].
  %%CITATION = JHEPA,0904,043;%%
  %[arXiv:0901.1145 [hep-ph]].
  %%CITATION = JHEPA,0904,114;%%
  %%CITATION = JHEPA,0507,065;%%
 S.~P.~Martin,
  %``Non-universal gaugino masses from non-singlet F-terms in non-minimal
  %unified models,''
  arXiv:0903.3568 [hep-ph];
  %%CITATION = ARXIV:0903.3568;%%
 D.~Feldman, Z.~Liu and P.~Nath,
  %``Gluino NLSP, Dark Matter via Gluino Coannihilation, and LHC Signatures,''
  arXiv:0905.1148 [hep-ph];
  %%CITATION = ARXIV:0905.1148;%%
  M.~Holmes and B.~D.~Nelson,
  %``Dark Matter Prospects in Deflected Mirage Mediation,''
  JCAP {\bf 0907}, 019 (2009);
%  [arXiv:0905.0674 [hep-ph]].
  %%CITATION = JCAPA,0907,019;%%
  S.~Bhattacharya, U.~Chattopadhyay, D.~Choudhury, D.~Das and B.~Mukhopadhyaya,
  %``Non-universal scalar mass scenario with Higgs funnel region of SUSY dark
  %matter: a signal-based analysis for the Large Hadron Collider,''
  arXiv:0907.3428 [hep-ph].
  %%CITATION = ARXIV:0907.3428;%%

  \bibitem{Hisano:2003ec}
  J.~Hisano, S.~Matsumoto and M.~M.~Nojiri,
  %``Explosive dark matter annihilation,''
  Phys.\ Rev.\ Lett.\  {\bf 92}, 031303 (2004).
%  [arXiv:hep-ph/0307216].
  %%CITATION = PRLTA,92,031303;%%

\bibitem{Cassel:2009pu}
  S.~Cassel, D.~M.~Ghilencea and G.~G.~Ross,
  %``Electroweak and Dark Matter Constraints on a Z' in Models with a Hidden
  %Valley,''
  arXiv:0903.1118 [hep-ph].
  %%CITATION = ARXIV:0903.1118;%%

\bibitem{Feldman:2006wd}
  D.~Feldman, B.~Kors and P.~Nath,
  %``Extra-weakly Interacting Dark Matter,''
  Phys.\ Rev.\  D {\bf 75}, 023503 (2007)
[arXiv:hep-ph/0610133];
  P.~Anastasopoulos, F.~Fucito, A.~Lionetto, G.~Pradisi, A.~Racioppi and Y.~S.~Stanev,
  %``Minimal Anomalous U(1) -prime Extension of the MSSM,''
  Phys.\ Rev.\  D {\bf 78}, 085014 (2008)
  %[arXiv:0804.1156 [hep-th]].
  %%CITATION = PHRVA,D78,085014;%%
  %%CITATION = PHLTA,B671,87;%%
  %%CITATION = PHRVA,D75,023503;%%
 F.~Fucito, A.~Lionetto, A.~Mammarella and A.~Racioppi,
  %``Axino Dark Matter in Anomalous U(1)' Models,''
  arXiv:0811.1953 [hep-ph];
  %%CITATION = ARXIV:0811.1953;%%
  C.~Coriano, M.~Guzzi, N.~Irges and A.~Mariano,
  %``Axion and Neutralinos from Supersymmetric Extensions of the Standard Model
  %with anomalous U(1)'s,''
  Phys.\ Lett.\  B {\bf 671} (2009) 87.
  %[arXiv:0811.0117 [hep-ph]].

\bibitem{Profumo:2006bx}
  S.~Profumo and A.~Provenza,
  %``Increasing the Neutralino Relic Abundance with Slepton Coannihilations:
  %Consequences for Indirect Dark Matter Detection,''
  JCAP {\bf 0612}, 019 (2006);[arXiv:hep-ph/0609290].
  %%CITATION = JCAPA,0612,019;%%

\bibitem{Griest:1990kh}
  K.~Griest and D.~Seckel,
  %``Three exceptions in the calculation of relic abundances,''
  Phys.\ Rev.\  D {\bf 43}, 3191 (1991).
  %%CITATION = PHRVA,D43,3191;%%

\bibitem{Gondolo:2004sc}
  P.~Gondolo, J.~Edsjo, P.~Ullio, L.~Bergstrom, M.~Schelke and E.~A.~Baltz,
  %``DarkSUSY: Computing supersymmetric dark matter properties numerically,''
  JCAP {\bf 0407}, 008 (2004);
  %%CITATION = JCAPA,0407,008;%%
  G.~Belanger, F.~Boudjema, A.~Pukhov and A.~Semenov,
  %``Dark matter direct detection rate in a generic model with micrOMEGAs2.1,''
  Comput.\ Phys.\ Commun.\  {\bf 180}, 747 (2009).
  %%CITATION = CPHCB,180,747;%

\bibitem{Kors:2004dx}
  B.~Kors and P.~Nath,
  %``A Stueckelberg extension of the standard model,''
  Phys.\ Lett.\  B {\bf 586}, 366 (2004);
%  [arXiv:hep-ph/0402047];
  %%CITATION = PHLTA,B586,366;%
  %``A supersymmetric Stueckelberg U(1) extension of the MSSM,''
  JHEP {\bf 0412}, 005 (2004);
%  [arXiv:hep-ph/0406167];
  %%CITATION = JHEPA,0412,005;%%,
  %``Aspects of the Stueckelberg extension,''
  JHEP {\bf 0507}, 069 (2005);
%  [arXiv:hep-ph/0503208];
  %%CITATION = JHEPA,0507,069;%%
 D.~Feldman, Z.~Liu and P.~Nath,
  %``The Stueckelberg $Z$ Prime at the LHC: Discovery Potential, Signature
  %Spaces and Model Discrimination,''
  JHEP {\bf 0611}, 007 (2006);
 % [arXiv:hep-ph/0606294].
  %%CITATION = JHEPA,0611,007;%%
 K.~Cheung and T.~C.~Yuan,
  %``Hidden fermion as milli-charged dark matter in Stueckelberg Z' model,''
  JHEP {\bf 0703}, 120 (2007);arXiv:0710.2005 [hep-ph];
   D.~Feldman, Z.~Liu and P.~Nath,
  %``The Stueckelberg Extension and Milli Weak and Milli Charge Dark Matter,''
  AIP Conf.\ Proc.\  {\bf 939}, 50 (2007).
 % [arXiv:0705.2924 [hep-ph]].
  %%CITATION = APCPC,939,50;%%
  %[arXiv:0712.2041 [hep-ph]].
  %%CITATION = JHEPA,0807,008;%%

\bibitem{FLNSt}
D.~Feldman, Z.~Liu and P.~Nath,
  %``Probing a Very Narrow $Z'$ Boson with CDF and D0 Data,''
  Phys.\ Rev.\ Lett.\  {\bf 97} (2006) 021801; See also
   J.~Kumar and J.~D.~Wells,
  %``LHC and ILC probes of hidden-sector gauge bosons,''
  Phys.\ Rev.\  D {\bf 74}, 115017 (2006)
 % [arXiv:hep-ph/0606183].
  W.~F.~Chang, J.~N.~Ng and J.~M.~S.~Wu,
  %``A Very Narrow Shadow Extra Z-boson at Colliders,''
  Phys.\ Rev.\  D {\bf 74}, 095005 (2006);
   S.~Gopalakrishna, S.~J.~Lee and J.~D.~Wells,
  %``Dark matter and Higgs boson collider implications of fermions in an
  %abelian-gauged hidden sector,''
  arXiv:0904.2007 [hep-ph].
  %%CITATION = ARXIV:0904.2007;%%
%

  \bibitem{Hisano:2005ec}
  J.~Hisano, S.~Matsumoto, O.~Saito and M.~Senami,
  %``Heavy Wino-like neutralino dark matter annihilation into antiparticles,''
  Phys.\ Rev.\  D {\bf 73} (2006) 055004.
  %[arXiv:hep-ph/0511118].
  %%CITATION = PHRVA,D73,055004;%%

  \bibitem{Delahaye:2007fr}
  T.~Delahaye, R.~Lineros, F.~Donato, N.~Fornengo and P.~Salati,
  %``Positrons from dark matter annihilation in the galactic halo: theoretical
  %uncertainties,''
  Phys.\ Rev.\  D {\bf 77}, 063527 (2008);
%  [arXiv:0712.2312 [astro-ph]].
%  %%CITATION = PHRVA,D77,063527;%%
  P.~Brun, G.~Bertone, J.~Lavalle, P.~Salati, R.~Taillet,
  %``Antiproton and Positron Signal Enhancement in Dark Matter Mini-Spikes
  %Scenarios,''
  Phys.\ Rev.\  D {\bf 76}, 083506 (2007).
  %[arXiv:0704.2543 [astro-ph]].
  %%CITATION = PHRVA,D76,083506;%%
%\cite{Adriani:2008zr}

\bibitem{Kamionkowski:2008vw}
  M.~Kamionkowski and S.~M.~Koushiappas,
  %``Galactic Substructure and Direct Detection of Dark Matter,''
  Phys.\ Rev.\  D {\bf 77}, 103509 (2008).
  %[arXiv:0801.3269 [astro-ph]].
  %%CITATION = PHRVA,D77,103509;%%

\bibitem{Longair:1994wu}
  M.~S.~Longair,
  %``High-energy astrophysics. Vol. 2: Stars, the galaxy and the interstellar
  %medium,''
%\href{http://www.slac.stanford.edu/spires/find/hep/www?irn=3524353}{SPIRES entry}
{\it  Cambridge, UK: Univ. Pr. (1994)}.

\bibitem{Navarro:1996gjMoore:1999gc}
  J.~F.~Navarro, C.~S.~Frenk and S.~D.~M.~White,
 % ``A Universal Density Profile from Hierarchical Clustering,''
  Astrophys.\ J.\  {\bf 490}, 493 (1997);
  B.~Moore et.al,
  %, T.~R.~Quinn, F.~Governato, J.~Stadel, G.~Lake,
  %``Cold collapse and the core catastrophe,''
  Mon.\ Not.\ Roy.\ Astron.\ Soc.\  {\bf 310}, 1147 (1999).
  %[arXiv:astro-ph/9903164].
  %%CITATION = MNRAA,310,1147;%%

  \bibitem{Corcella:2000bw}
  G.~Corcella {\it et al.},
  %``HERWIG 6.5: an event generator for Hadron Emission Reactions With
  %Interfering Gluons (including supersymmetric processes),''
  JHEP {\bf 0101}, 010 (2001).
%  [arXiv:hep-ph/0011363].
  %%CITATION = JHEPA,0101,010;%%

\bibitem{Moskalenko:1997gh}
  I.~V.~Moskalenko and A.~W.~Strong,
  %``Production and propagation of cosmic-ray positrons and electrons,''
  Astrophys.\ J.\  {\bf 493}, 694 (1998);
 % [arXiv:astro-ph/9710124].
  %%CITATION = ASJOA,493,694;%%
  E.~A.~Baltz and J.~Edsjo,
  %``Positron Propagation and Fluxes from Neutralino Annihilation in the Halo,''
  Phys.\ Rev.\  D {\bf 59}, 023511 (1998).
%  [arXiv:astro-ph/9808243].
  %%CITATION = PHRVA,D59,023511;%%

\bibitem{Bottino:1994xs}
  A.~Bottino, C.~Favero, N.~Fornengo and G.~Mignola,
  %``Amount of anti-protons in cosmic rays due to halo neutralino
  %annihilation,''
  Astropart.\ Phys.\  {\bf 3}, 77 (1995).
%  [arXiv:hep-ph/9408392].
  %%CITATION = APHYE,3,77;%%

\bibitem{Maurin:2001sj}
  D.~Maurin, F.~Donato, R.~Taillet and P.~Salati,
  %``Cosmic Rays below Z=30 in a diffusion model: new constraints on propagation
  %parameters,''
  Astrophys.\ J.\  {\bf 555}, 585 (2001).
%  [arXiv:astro-ph/0101231].
  %%CITATION = ASJOA,555,585;%%

\bibitem{Bergstrom:1999jc}
    L.~Bergstrom, J.~Edsjo and P.~Ullio,
  %``Cosmic antiprotons as a probe for supersymmetric dark matter?,''
  Astrophys.\ J.\  {\bf 526}, 215 (1999).
%  [arXiv:astro-ph/9902012].
  %%CITATION = ASJOA,526,215;%%

\bibitem{Bringmann:2006im}
  T.~Bringmann and P.~Salati,
  %``The galactic antiproton spectrum at high energies: background   expectation
  %vs. exotic contributions,''
  Phys.\ Rev.\  D {\bf 75}, 083006 (2007).
%  [arXiv:astro-ph/0612514].
  %%CITATION = PHRVA,D75,083006;%%

\bibitem{Bruno}
A.~Bruno, 
%``Cosmic ray antiprotons measured
%in the PAMELA experiment''; 
PhD Thesis, Universita Degli Studi Di Bari.
%under R.~Bellotti, F.~S.~Cafagna.

\bibitem{Ricci}
M.~Ricci, Invited Talk at SUSY 09, Northeastern University, Boston.

\bibitem{Orito:1999re}
 S.~Orito {\it et al.}  [BESS Collaboration],
 %``Precision measurement of cosmic-ray antiproton spectrum,''
 Phys.\ Rev.\ Lett.\  {\bf 84}, 1078 (2000);
%  [arXiv:astro-ph/9906426].
 %%CITATION = PRLTA,84,1078;%%
See also:
 T.~Maeno {\it et al.}  [BESS Collaboration],
 %``Successive measurements of cosmic-ray antiproton spectrum in a positive
 %phase of the solar cycle,''
 Astropart.\ Phys.\  {\bf 16}, 121 (2001);
 %[arXiv:astro-ph/0010381].
 %%CITATION = APHYE,16,121;%%
 M.~Boezio {\it et al.}  [WiZard/CAPRICE Collaboration],
 %``The cosmic-ray anti-proton flux between 3-GeV and 49-GeV,''
 Astrophys.\ J.\  {\bf 561}, 787 (2001).
%  [arXiv:astro-ph/0103513].
 %%CITATION = ASJOA,561,787;%%

\bibitem{Bergstrom:1997fj}
  L.~Bergstrom, P.~Ullio and J.~H.~Buckley,
  %``Observability of gamma rays from dark matter neutralino annihilations  in
  %the Milky Way halo,''
  Astropart.\ Phys.\  {\bf 9}, 137 (1998).
%  [arXiv:astro-ph/9712318].
  %%CITATION = APHYE,9,137;%%

\bibitem{Cirelli:2009vg}
  P.~Meade, M.~Papucci, A.~Strumia and T.~Volansky,
  %``Dark Matter Interpretations of the Electron/Positron Excesses after
  %FERMI,''
  arXiv:0905.0480 [hep-ph];
  M.~Cirelli and P.~Panci,
  %``Inverse Compton constraints on the Dark Matter e+e- excesses,''
  arXiv:0904.3830 [astro-ph.CO].
  %%CITATION = ARXIV:0904.3830;%%

\bibitem{Berezinsky:1994wva}
  V.~Berezinsky, A.~Bottino and G.~Mignola,
  %``High-energy gamma radiation from the galactic center due to neutralino
  %annihilation,''
  Phys.\ Lett.\  B {\bf 325}, 136 (1994).
%  [arXiv:hep-ph/9402215].
  %%CITATION = PHLTA,B325,136;%%

\bibitem{Fornengo:2004kj}
  N.~Fornengo, L.~Pieri and S.~Scopel,
  %``Neutralino annihilation into gamma-rays in the Milky Way and in  external
  %galaxies,''
  Phys.\ Rev.\  D {\bf 70}, 103529 (2004).
%  [arXiv:hep-ph/0407342].
  %%CITATION = PHRVA,D70,103529;%%

\bibitem{Strong:2004de}
  A.~W.~Strong, I.~V.~Moskalenko and O.~Reimer,
  %``Diffuse Galactic continuum gamma rays. A model compatible with EGRET data
  %and cosmic-ray measurements,''
  Astrophys.\ J.\  {\bf 613}, 962 (2004).
%  [arXiv:astro-ph/0406254].
  %%CITATION = ASJOA,613,962;%%
  %\cite{Cirelli:2009vg}

\bibitem{Bergstrom:2005ss}
  L.~Bergstrom, T.~Bringmann, M.~Eriksson and M.~Gustafsson,
  %``Gamma rays from heavy neutralino dark matter,''
  Phys.\ Rev.\ Lett.\  {\bf 95}, 241301 (2005).
%  [arXiv:hep-ph/0507229].
  %%CITATION = PRLTA,95,241301;%%

\bibitem{Hunger:1997we}
  S.~D.~Hunter {\it et al.},
  %``EGRET observations of the diffuse gamma-ray emission from the galactic
  %plane,''
  Astrophys.\ J.\  {\bf 481}, 205 (1997).
  %%CITATION = ASJOA,481,205;%%

\bibitem{gig}
N.~Giglietto, Les Rencontes de Physique, de la Vallee D'Aoste;
La Thuile 2009.

\bibitem{winer}
B.~Winer, Plenary Talk at SUSY 09, Northeastern University, Boston;
  T.~A.~Porter and f.~t.~F.~Collaboration,
  %``Fermi LAT Measurements of the Diffuse Gamma-Ray Emission at Intermediate
  %Galactic Latitudes,''
  arXiv:0907.0294 [astro-ph.HE].
  %%CITATION = ARXIV:0907.0294;%%


\bibitem{Bergstrom:1997fh}
  L.~Bergstrom and P.~Ullio,
  %``Full one-loop calculation of neutralino annihilation into two photons,''
  Nucl.\ Phys.\  B {\bf 504}, 27 (1997);
  %[arXiv:hep-ph/9706232].
  %%CITATION = NUPHA,B504,27;%%
  Z.~Bern, P.~Gondolo and M.~Perelstein,
  %``Neutralino annihilation into two photons,''
  Phys.\ Lett.\  B {\bf 411}, 86 (1997).
  %%[arXiv:hep-ph/9706538].
  %%CITATION = PHLTA,B411,86;%%


\bibitem{ben}
  B.~C.~Allanach,
  %``SOFTSUSY: A C++ program for calculating supersymmetric spectra,''
  Comput.\ Phys.\ Commun.\  {\bf 143}, 305 (2002)
%  [arXiv:hep-ph/0104145];
  Present version 3.02 (2009).
  %%CITATION = CPHCB,143,305;%%


%\cite{Hisano:2004pv}
\bibitem{Hisano:2004pv}
  J.~Hisano, S.~Matsumoto, M.~M.~Nojiri and O.~Saito,
  %``Direct detection of the Wino- and Higgsino-like neutralino dark matters  at
  %one-loop level,''
  Phys.\ Rev.\  D {\bf 71}, 015007 (2005).
%  [arXiv:hep-ph/0407168].
  %%CITATION = PHRVA,D71,015007;%%



\bibitem{pythiapgs}
  P.~Skands {\it et al.},
  %``SUSY Les Houches accord: Interfacing SUSY spectrum calculators, decay
  %packages, and event generators,''
  JHEP {\bf 0407}, 036 (2004); PGS-4, J.~Conway et. al.
%  [arXiv:hep-ph/0311123].
  %%CITATION = JHEPA,0407,036;%%

\bibitem{Ball:2007zza}
  G.~L.~Bayatian {\it et al.}  [CMS Collaboration],
  %``CMS technical design report, volume II: Physics performance,''
  J.\ Phys.\ G {\bf 34}, 995 (2007).
  %%CITATION = JPHGB,G34,995;%%

\bibitem{Baer:1995nq}
  H.~Baer, C.~h.~Chen, F.~Paige and X.~Tata,
  %``Signals for minimal supergravity at the CERN large hadron collider: Multi -
  %jet plus missing energy channel,''
  Phys.\ Rev.\  D {\bf 52}, 2746 (1995);
%  [arXiv:hep-ph/9503271].
  %%CITATION = PHRVA,D52,2746;%%
H.~Baer, C.~h.~Chen, F.~Paige and X.~Tata,
 %``Signals for Minimal Supergravity at the CERN Large Hadron Collider II:
 %Multilepton Channels,''
 Phys.\ Rev.\  D {\bf 53}, 6241 (1996)
 [arXiv:hep-ph/9512383].
 %%CITATION = PHRVA,D53,6241;%%

\bibitem{Chattopadhyay:2006xb}
  U.~Chattopadhyay, D.~Das, P.~Konar and D.~P.~Roy,
  %``Looking for a heavy wino LSP in collider and dark matter experiments,''
  Phys.\ Rev.\  D {\bf 75}, 073014 (2007).
%  [arXiv:hep-ph/0610077].
  %%CITATION = PHRVA,D75,073014;%%

\bibitem{Paige:1999ui}
  F.~E.~Paige and J.~D.~Wells,
%  ``Anomaly mediated SUSY breaking at the LHC,''
  arXiv:hep-ph/0001249.
  %%CITATION = HEP-PH/0001249;%%

\bibitem{Baer:2000bs}
  H.~Baer, J.~K.~Mizukoshi and X.~Tata,
  %``Reach of the CERN LHC for the minimal anomaly-mediated SUSY breaking
  %model,''
  Phys.\ Lett.\  B {\bf 488}, 367 (2000).
 % [arXiv:hep-ph/0007073].
  %%CITATION = PHLTA,B488,367;%%

\bibitem{Giudice:1998xp}
  L.~Randall and R.~Sundrum,
  %``Out of this world supersymmetry breaking,''
  Nucl.\ Phys.\  B {\bf 557}, 79 (1999);
  [arXiv:hep-th/9810155].
  %%%CITATION = NUPHA,B557,79;%%
  G.~F.~Giudice, M.~A.~Luty, H.~Murayama and R.~Rattazzi,
  %``Gaugino Mass without Singlets,''
  JHEP {\bf 9812}, 027 (1998);
 [arXiv:hep-ph/9810442].
  %%CITATION = JHEPA,9812,027;%%

\bibitem{Gaillard:1999yb}
 M.~K.~Gaillard, B.~D.~Nelson and Y.~Y.~Wu,
 %``Gaugino masses in modular invariant supergravity,''
 Phys.\ Lett.\  B {\bf 459} (1999) 549
 [arXiv:hep-th/9905122].
 %%CITATION = PHLTA,B459,549;%%

\bibitem{Bagger:1999rd}
 J.~A.~Bagger, T.~Moroi and E.~Poppitz,
 %``Anomaly mediation in supergravity theories,''
 JHEP {\bf 0004}, 009 (2000)
 [arXiv:hep-th/9911029].
 %%CITATION = JHEPA,0004,009;%%

\bibitem{Chen:1995yu}
  C.~H.~Chen, M.~Drees and J.~F.~Gunion,
  %``Searching for Invisible and Almost Invisible Particles at $e~+e~-$
  %Colliders,''
  Phys.\ Rev.\ Lett.\  {\bf 76}, 2002 (1996).
%  [arXiv:hep-ph/9512230].
  %%CITATION = PRLTA,76,2002;%%

\bibitem{Feng:1999fu}
  J.~L.~Feng, T.~Moroi, L.~Randall, M.~Strassler and S.~f.~Su,
  %``Discovering supersymmetry at the Tevatron in Wino LSP scenarios,''
  Phys.\ Rev.\ Lett.\  {\bf 83}, 1731 (1999).
%  [arXiv:hep-ph/9904250].
  %%CITATION = PRLTA,83,1731;%%

\bibitem{Gherghetta:1999sw}
  T.~Gherghetta, G.~F.~Giudice and J.~D.~Wells,
  %``Phenomenological consequences of supersymmetry with anomaly-induced
  %masses,''
  Nucl.\ Phys.\  B {\bf 559}, 27 (1999).
%  [arXiv:hep-ph/9904378].
  %%CITATION = NUPHA,B559,27;%%

\bibitem{Barr:2002ex}
  A.~J.~Barr, C.~G.~Lester, M.~A.~Parker, B.~C.~Allanach and P.~Richardson,
  %``Discovering anomaly-mediated supersymmetry at the LHC,''
  JHEP {\bf 0303}, 045 (2003).
%  [arXiv:hep-ph/0208214].
  %%CITATION = JHEPA,0303,045;%%

\bibitem{Kitano:2006gv}
  R.~Kitano and Y.~Nomura,
  %``Supersymmetry, naturalness, and signatures at the LHC,''
  Phys.\ Rev.\  D {\bf 73}, 095004 (2006).
%  [arXiv:hep-ph/0602096].
  %%CITATION = PHRVA,D73,095004;%%

\bibitem{Hinchliffe:1996iu}
  I.~Hinchliffe, F.~E.~Paige, M.~D.~Shapiro, J.~Soderqvist and W.~Yao,
  %``Precision SUSY measurements at CERN LHC,''
  Phys.\ Rev.\  D {\bf 55}, 5520 (1997).
%  [arXiv:hep-ph/9610544].
  %%CITATION = PHRVA,D55,5520;%%

\bibitem{fermilat}
  D.~Grasso {\it et al.}  [FERMI-LAT Collaboration],
  %``On possible interpretations of the high energy electron-positron spectrum
  %measured by the Fermi Large Area Telescope,''
  arXiv:0905.0636 [astro-ph.HE];
%\cite{Bergstrom:2009fa}
  L.~Bergstrom, J.~Edsjo and G.~Zaharijas,
  %``Dark matter interpretation of recent electron and positron data,''
  arXiv:0905.0333 [astro-ph.HE].
  %%CITATION = ARXIV:0905.0333;%%




\end{thebibliography}
\end{document}